\documentclass{article}
\usepackage[utf8]{inputenc}
\usepackage{comment}

\usepackage[sort]{natbib}
\usepackage{hyperref}
\usepackage{authblk}

\usepackage{graphicx}
\usepackage{wrapfig}
\usepackage{subcaption}
\usepackage{float}
\usepackage[font=small,labelfont=bf]{caption}
\usepackage[normalem]{ulem}

\usepackage{placeins}

\usepackage{amsmath}
\usepackage{amsfonts}
\usepackage{amssymb}
\usepackage[capitalize]{cleveref}

\usepackage{pdfpages}

\usepackage[most]{tcolorbox}
\newtcolorbox[auto counter]{mathbox}[2][]
{
  center,
  breakable,
  width = 1.09\linewidth,
  colframe = gray!25,
  colback  = gray!10,
  coltitle = black,  
  title    = \textbf{Box \thetcbcounter. #2},
  #1,
}

\usepackage[draft]{todonotes}   

\title{Sources of Richness and Ineffability for Phenomenally Conscious States}

\author[1,2]{Xu Ji\footnote{Equal contribution. Correspondence to \texttt{\{xu.ji,eric.elmoznino\}@mila.quebec
}}}
\author[1,2]{Eric Elmoznino$^*$}
\author[2]{\authorcr George Deane}
\author[3]{Axel Constant}
\author[1,2]{Guillaume Dumas}
\author[1,2]{Guillaume Lajoie}
\author[2]{Jonathan Simon}
\author[1,2,4]{Yoshua Bengio}

\affil[1]{Mila}
\affil[2]{University of Montreal}
\affil[3]{University of Sussex}
\affil[4]{CIFAR Fellow}

\date{\today}

\begin{document}

\maketitle

\begin{abstract}

Conscious states---states that there is something it is like to be in---seem both rich or full of detail, and ineffable or hard to fully describe or recall. The problem of ineffability, in particular, is a longstanding issue in philosophy that partly motivates the explanatory gap: the belief that consciousness cannot be reduced to underlying physical processes. Here, we provide an information theoretic dynamical systems perspective on the richness and ineffability of consciousness. In our framework, the richness of conscious experience corresponds to the amount of information in a conscious state and ineffability corresponds to the amount of information lost at different stages of processing. We describe how attractor dynamics in working memory would induce impoverished recollections of our original experiences, how the discrete symbolic nature of language is insufficient for describing the rich and high-dimensional structure of experiences, and how similarity in the cognitive function of two individuals relates to improved communicability of their experiences to each other. While our model may not settle all questions relating to the explanatory gap, it makes progress toward a fully physicalist explanation of the richness and ineffability of conscious experience---two important aspects that seem to be part of what makes qualitative character so puzzling.
  
\end{abstract}

\clearpage
\tableofcontents
\clearpage

\section{Introduction}
\label{sec:intro}

Conscious states---states that there is something it is like to be in \citep{nagel1974like}---present many apparent contradictions. On the one hand, every time we have a thought, look out at the world, or feel an emotion, we have a rich experience that seems impossible to fully describe. At the same time, conscious experiences are conceptualizable, with similar properties across individuals, and can often be communicated with a degree of fidelity.

This paper provides an information theoretic dynamical systems perspective on how and why consciousness may appear to us the way it does, namely as both \emph{rich} or full of detail, and \emph{ineffable} or hard to fully describe or recall---in other words, why it seems that an experience is ``worth a thousand words''. In addition, a dynamical systems model for consciousness offers an explanation for why much of the conscious content that \emph{is} reportable has a discrete nature that can be expressed with words. Our key contention is that these aspects of consciousness are implicated by a dynamical systems model of neural processing, in particular by ``attractors'': patterns of joint neural activity that remain relatively stable over short timescales and yield a discrete partition over neural states. 
Importantly, interpreting cognitive processing through the lenses of dynamical systems and information theory will give us the ability to reason about richness, ineffability, and communicability in general terms, without relying on implementation details of the neural processes that may give rise to consciousness. Broadly, the suggestion is that the rather abstract level of explanation afforded by information theory is the commensurate level of explanation for some key questions about richness and ineffability.

By ``consciousness'' we mean phenomenal consciousness, i.e. the felt or subjective quality of experience. A state is phenomenally conscious when, in the words of \citet{nagel1974like}, there is \textit{something it is like} to be in that state. Phenomenal consciousness is the form of consciousness that gives rise to what Joseph Levine calls the ``explanatory gap'' \citep{levine1993leaving} and what David Chalmers calls the ``hard problem of consciousness'' \citep{chalmers1996conscious}: the problem of showing that phenomenal consciousness can be explained in terms of, or reduced to, underlying physical processes. The explanatory gap is one of the central problems in the philosophy of mind, and it relies heavily on the intuition that ``physicalist theories leave out [phenomenal consciousness] in the epistemological sense, because they reveal our inability to explain qualitative character in terms of the physical properties of sensory states'' \citep{levine1993leaving}.

Here, we address one aspect of this problem by developing a structural/mechanistic explanation of the richness and ineffability of conscious experience, one that is given entirely in terms of information processing in a dynamical system such as the brain. Our model assumes that conscious experiences are derived from neural processes according to known physical laws, and can therefore be understood using the standard methods of cognitive neuroscience. While our model may not settle all questions relating to the explanatory gap, it will make progress toward a fully physicalist explanation of the richness and ineffability of conscious experience---two important aspects that seem to be part of what makes qualitative character so puzzling. Richness and ineffability figure in several important live debates about consciousness in the philosophical literature. Here we summarize two: the illusionism debate and the overflow debate.

Illusionists argue that consciousness is an illusion, while realists deny this \citep{frankish2016illusionism}. Illusionists generally argue that our expectations for consciousness are too high: that the job of describing a conscious experience is too demanding for any physical process to fulfill, and that (rather than rejecting physicalism) we should conclude that there is no such thing as consciousness (or at least, make do with a diminished conception of it) \citep{dennett1993consciousness,graziano2020toward,humphrey2020invention}. Daniel Dennett famously lists ineffability as one of the hard-to-fulfill conditions that should lead one to illusionism: the prospect that conscious contents somehow escape our attempts to fully describe them is, for Dennett, a sign that consciousness is chimerical \citep{dennett1993consciousness}. Notably, illusionists acknowledge that something gives rise to the relevant illusions: there must be an explanation of why it seems plausible to us, on introspection, that we are the subjects of (ineffable) conscious states. Qualia realists, in contrast, see conscious experience as the subjective viewpoint from which all else is observed or known, and therefore consider it to be an explanandum that cannot be discarded \citep{tononi1998consciousness,chalmers2010character,descartes1986}.

The overflow debate is between those who hold that consciousness is indeed rich and ineffable, and those who deny it (while still maintaining that consciousness exists). Richness is a relative term, and one contender for a reference object that justifies the characterization of consciousness as rich is the accessible content of working memory.
Empirically there appears to be a clear bandwidth limitation on the latter \citep{sperling1960information, MillerBuschman2015, Cohen2016-COHWIT-3}, which is what makes it difficult, for example, to remember all of the names of the people you meet at a party or all of the digits of a phone number. Proponents of overflow say that consciousness is considerably richer than this sort of working memory and includes ineffable content unavailable for report \citep{block2007overflow,bronfman2019impoverished,lamme2007sue,vandenbroucke2012non}, while the staunchest opponents of overflow will maintain that consciousness is no richer than the bandwidth-restricted content of working memory, generally because they take consciousness to just \emph{be} working memory or a supporting system for it \citep{ward2018downgraded,phillips2016no,naccache2018and,cohen2011consciousness}.

We thus have two important debates where those on both sides may benefit from a formal model of ineffability: illusionists and realists who deny overflow may benefit from a general model of why it seems to us that we are the subjects of rich and ineffable experiences, while realists who accept overflow may benefit from a characterization of how it emerges.

The aim of this paper is to propose and justify a formal description of how neural dynamics could give rise to the ordinary sense of richness and ineffability in the brain. Our key contributions are summarized as follows:

\begin{itemize}
    \item We relate the philosophical notions of richness and ineffability to the computational notion of information. 
    Assuming that brain dynamics are cast as information processing functions, we contend that the richness of conscious experience can be interpreted as the amount of information in conscious state, and ineffability as the amount of information lost in processing. 

    \item 
    Attractor dynamics are empirically ubiquitous in neural activity across cortical regions and have been proposed as a computational model for working memory \citep{khona2022attractor,rolls2010attractor}, while prominent models of consciousness argue that conscious experience is a projection of working memory states \citep{baars2005,dehaene2001towards}.
    We connect these theories by contending that significant information loss induced by attractor dynamics offers an account for the significant ineffability of conscious experience.


    \item By considering information at multiple stages during inter-personal communication, we show how different point-to-point pathways of information loss arise during cognitive processing, going beyond the specific case of ineffability of conscious experience at verbal report.

    \item Using Kolmogorov information theory \citep{kolmogorov1965three} we prove a formal result that connects cognitive dissimilarity between individuals with increasing ineffability of conscious experience. 
    This highlights the difference between cognitive dissimilarity and knowledge inadequacy, shedding light on the philosophical conundrum of what color scientist Mary learns when leaving her black and white room \citep{jackson1986}. 

    \item 
    Since information loss is a function of neural states, it can be approximately computed by cognitive processing, providing a mechanistic justification for the report of ineffability, or the contention that consciously inaccessible rich representations exist \citep{sperling1960information}. 
    
\end{itemize}



Several existing works argue that attractor dynamics have the right functional characteristics to serve as a computational model for consciousness \citep{colagrosso2004theories,mozer2009attractor,mathis1994computational,mathis2019conscious,grossberg1999link,rumelhart1986sequential} but do not examine how information loss arising from such dynamics relate to the rich and ineffable aspects of conscious experience.
Instead of defining ``access'' as triggering correct behavior on a per-experience basis \citep{colagrosso2004theories}, we contend that there is a natural correspondence between access and preservation of information, which allows for quantification using mutual information and analysis by applying information theoretic reasoning to the abstract computation graph. 
We utilize a minimal computational model without relying on  implementation details of neural processing functions to maximize the generality of arguments.
Casting ineffability as information loss allows us to reason about the ineffability of conscious experience from the computation graph without depending on the exact definition of conscious experience. 

The paper is structured as follows. In \cref{sec:neural_dynamics}, we introduce key concepts on computation and neural dynamics, in particular the role of attractor states that can be used for computations involving short-term memory and have a dual discrete and high-dimensional nature. 
We present our dynamical systems model of conscious experience in \cref{sec:consious_dynamics}, beginning with \cref{sec:motivating_conscious_dynamics} which motivates the use of attractor dynamics for modeling conscious processing using prior arguments from the literature that are independent of our own, including evidence for the Global Workspace Theory \citep{baars1993,baars2005,dehaene1998}.
\cref{sec:richness_and_ineffability} formalizes the notions of richness and ineffability using both Shannon information theory \citep{shannon1948mathematical} and Kolmogorov complexity \citep{kolmogorov1965three}, which play a central role in making our later arguments precise. 
Core contributions are presented in \cref{sec:intra,sec:inter}, which discuss various sources of ineffability in conscious experience and explain the conditions under which these experiences can be partially communicated to others. We then briefly discuss the implications of our model on the debate surrounding `phenomenal' vs. `access' consciousness \citep{block1995}, before concluding with a high-level discussion in \cref{sec:discussion}.
\section{Preliminaries: Computation through neural dynamics}
\label{sec:neural_dynamics} 

In Section~\ref{sec:consious_dynamics}, we will argue that we can account for the richness and ineffability of experience by modeling conscious states as neural trajectories in a high-dimensional dynamical system with attractors. To do so, we will now provide a brief overview of the essential concepts needed to understand the model.

First, we will introduce the notion of a neural activation space, in which temporally evolving states of neural activity follow trajectories governed by recurrent dynamics in the brain. Next, we will explain how state attractors, which are emergent properties of dynamical systems, can allow neural networks to solve computational problems that require some form of persistent memory. Along the way, we will highlight key examples from the computational neuroscience literature where this dynamical systems framework was used to explain how populations of neurons solve perceptual and cognitive tasks.

\subsection{Neural activation state space}

\begin{figure}[ht]
    \centering
    \includegraphics[width=\linewidth]{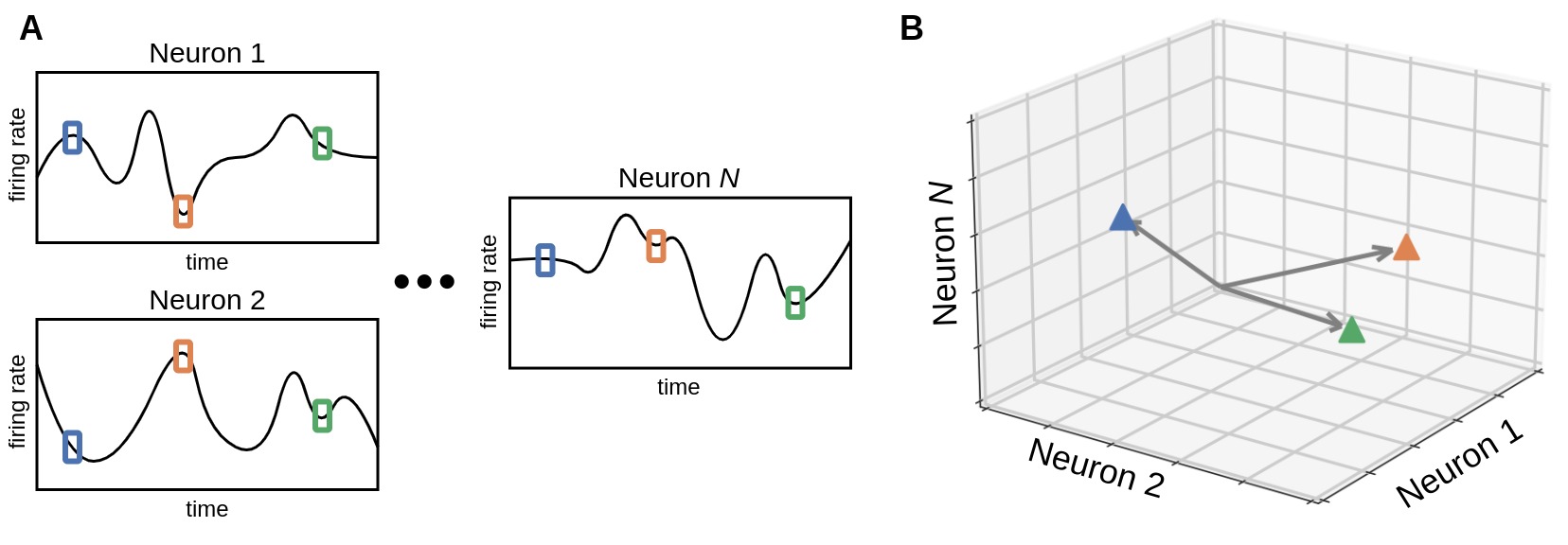}
    \caption{\textbf{Visualization of neural state space.} \textbf{A.} The activity trace for multiple neurons, where activity can be quantified in several different ways (e.g., firing or not, firing rate over some time window, membrane voltage, etc.). Colored boxes denote joint activity patterns across all neurons at specific timepoints. \textit{B.} At any particular timepoint, the joint activity pattern across $N$ different neurons can be expressed as a vector in an $N$-dimensional state space.}
    \label{fig:neural_activation_space}
\end{figure}

At any given moment, every neuron in the brain has some level of activity, and this activity can be numerically quantified in several different ways (e.g., firing or not, firing rate over some time window, membrane voltage, etc.), which we illustrate in \cref{fig:neural_activation_space}a. Together, this instantaneous pattern of activity defines the brain's current \textit{state}, which may be compactly represented as a vector in an $N$-dimensional state space, where $N$ is the number of neurons in the brain (or in the subpopulation of interest). In such a representation, each index in this vector identifies a particular neuron, and the value of a particular index corresponds to that neuron's current level of activity (\cref{fig:neural_activation_space}b).
We reason at the level of neuronal activity for clarity, but strictly our framework makes no assumptions about the appropriate level of granularity: where these make direct contributions to cognitive information processing, other cells such as astrocytes or cell components such as dendrites may be state-space parameters in their own right \citep{Godfrey-Smith2016-GODMMA-6}.

A benefit of describing neural activity in this manner is that it allows us to draw on the mathematical framework of dynamical systems theory to reason about mental states. For example, we can now talk about what a pattern of neural activity \textit{represents} by projecting the state onto lower-dimensional subspaces that encode some meaningful feature. To explain this by example, it might be the case that when perceiving an object certain dimensions of the elicited state represent its color, others represent its shape, yet others represent its function, etc. In addition, given a probabilistic transition model for states that accounts for noise in neural activity and other sources of uncertainty, we can measure quantities such as the likelihood and information content of a state. We can also quantify the similarities between states according to some distance metric between their vectors.

\subsection{Neural dynamics}

While neural states can be used to represent an instantaneous pattern of activity, the brain is a complex dynamical system and must ultimately be understood in terms of how neural activity unfolds in time. The temporal evolution of neural activity---and any other dynamical system---is governed by two factors.

\begin{figure}[ht]
    \centering
    \includegraphics[width=\linewidth]{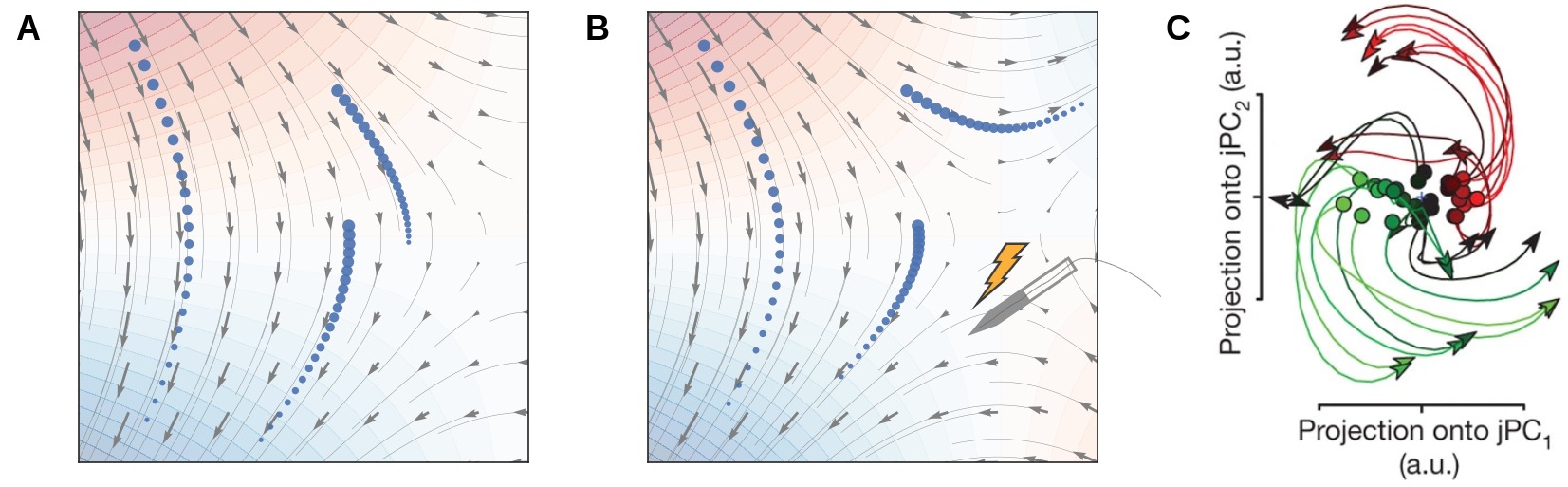}
    \caption{\textbf{Neural dynamics and trajectories in activation space.} \textbf{A.} A dynamical system whose behavior is depicted using vector fields and example trajectories. \textbf{B.} External inputs can modulate the behavior of a dynamical system (compare vector fields and trajectories with those in Panel A). \textbf{C.} An example of neural dynamics empirically observed in the primate motor cortex. As the neural dynamics are high-dimensional, jPCA was used to reduce their dimensionality for visualization. The figure was reproduced with permissions from \citet{churchland2012neural}.}
    \label{fig:dynamics}
\end{figure}

First, neurons in the brain have a large number of synapses that form recurrent loops. Recurrency means that even in the absence of any sensory input, brain states will evolve dynamically; the activity of one neuron at a particular time will influence the future activity of surrounding neurons, which may in turn influence the original neuron's activity at a later time in a causal loop. The dynamics governing these neural state trajectories are defined by the joint synaptic connectivity profile between all neurons in the brain. Any given connectivity profile results in a set of rules for how each state transitions to the next. This can be visually illustrated for the entire system using a \textit{vector field} as shown in \cref{fig:dynamics}a: each vector indicates how a state at that location would evolve in the next instant in the absence of noise, and where the size of the vector denotes the speed of the change. Intuitively, one can understand the dynamics of the system by starting off at an initial point in neural state space and tracing a trajectory that follows the vector field at each point in time. A different connectivity profile would yield different transition dynamics (i.e., a different vector field), and therefore the same initial neural state would follow a different trajectory.

Another factor that governs neural dynamics is the input to the system, which may itself evolve over time. The dynamics of a sub-population of neurons (e.g., a particular brain region) are modulated extrinsically by signals from surrounding neurons that synapse onto the population, including information from the stream of sensory signals entering the brain. Illustrated visually in \cref{fig:dynamics}b, this means that inputs warp the vector field that define transitions from the current state to the next, ultimately resulting in potentially very different trajectories from those that would have occurred given other inputs.

Much of the field of computational neuroscience is concerned with understanding neural population coding through the lens of dynamical systems, thanks to their rich theoretical underpinnings and the mechanistic models they provide \citep{favela2021dynamical}. Historically, this approach has been particularly fruitful in two systems: sensory integration \citep{burak2014spatial,zhang1996representation} and motor control \citep{churchland2012neural,michaels2016neural,shenoy2013cortical}. For example, \citet{churchland2012neural} recorded from a population of neurons in the primate motor cortex and found that they exhibited rotational dynamics during a simple reaching task (\cref{fig:dynamics}c). While this was initially surprising because the movement itself was not rhythmic, the authors proposed a theory that muscle activity is constructed from an oscillatory basis, which was later supported by additional experiments. The neural dynamics, then, can be understood as pattern generators that generate sequences of muscle activity optimized for producing natural movements.

Despite the success of this framework in sensory and motor domains, much less is understood about the dynamical underpinnings of higher-level cognition, although such dynamical systems are also implemented with neural substrates and would presumably share similar mechanisms. A contribution of our work is the application of dynamical systems to high-level conscious cognition and analysis of the implications for explaining the richness and ineffability of experience.

\subsection{State attractors}
\label{sec:attractors}

When neural dynamics are used to solve computational tasks, it is often the case that the solutions require some form of persistent memory, meaning that at least some projections of the neural activity must be self-sustaining. A dynamical system can implement this behavior by forming regions in its state space where states are drawn towards steady states (\cref{fig:attractors}a). These regions are called ``basins of attraction'' because any state trajectory that enters them would progress towards the steady state in the absence of noise or changes in external inputs and dynamics.
By steady states, we mean regions within the basins that deterministic trajectories eventually converge to. More generally, these sets of states are called ``attractors'' because neural activity trajectories that have reached the basin progress towards attractor states and remain there---approximately, in the presence of intrinsic noise in neural activity or changes in external input---until sufficient noise or external input activity nudge the state to escape the attractor basin. 
In general, dynamical systems can produce attractors that have complex and high-dimensional structure within the basin (e.g. manifold, fractal structure) and can exhibit their own internal dynamics, as in is the case of chaotic attractors (also called ``strange'').
Other common attractors contain fewer points, such as stable periodic orbits, or stable fixed points---single state points that do not change in time. 
In this section we will focus on fixed point attractors for simplicity, but arguments in subsequent sections apply to the general case of attractor subspaces. The important aspect of attractors for our purposes is that they are distinct and have non-overlapping basins of attraction.

Since trajectories that have converged to attractors have a tendency to remain there in the absence of strong external inputs, attractors can endow a dynamical system with a form of self-sustaining memory over short timescales that is useful for performing many computations essential to real-world tasks.
Attractor dynamics can also be used for efficient long-term memory, without the brain having to directly store the high-dimensional vectors of the attractors in state space. As we will explain in \cref{sec:discretization}, attractors are mutually exclusive and thus have a discrete structure; they can be identified with symbols (e.g., words) that label \textit{which} attractor the system is in without describing the attractor's location in state space. The system could thus store a concise symbol in long-term memory rather than a high-dimensional vector. Afterwards, the memory could be retrieved by using the symbol as an input `key' that drives the state to any location in the basin of the attractor, at which point the dynamics of the system will cause the trajectory to converge to the attractor. For example, to memorize an image of a face (represented by a high-dimensional vector) and associate it with a discrete entity like the name of a person, a learning process could update the parameters of the dynamical system, so that the image vector is an attractor state and system enters its basin of attraction when the name (or rather a neural code for it) is provided as an input. 

It is important to emphasize that the existence of these attractors and the particular properties they have (e.g., cardinality, location, shape) are purely functions of the internal dynamics of the system. Neural networks are therefore particularly well-suited for implementing diverse computations through dynamical systems since they are composed of simple units whose connectivity can be flexibly tuned to achieve many possible complex attractor configurations, with the capacity for universal function approximation in the limit of large networks \citep{schafer2007recurrent}.

A dynamical system can be modulated by external inputs, therefore the nature of its attractors can also be driven by contextual signals. In the human brain, for example, this context could include both external sensory input and the content of short- and long-term memory. In particular, the previous content of working memory (which is a part of short term memory) might have a strong influence, so that our thoughts form coherent sequences and so that we can alternate between mutually exclusive interpretations of the world that are compatible with the context (e.g., flipping between different interpretations of the \href{https://en.wikipedia.org/wiki/Necker_cube}{Necker cube}---an ambiguous 2D line drawing of a cube that can be in one of two possible 3D orientations).

As was summarized in review articles by \citet{rolls2010attractor} and \citet{khona2022attractor}, the framework of attractor dynamics has been used to mechanistically explain the neural computations underlying decision-making \citep{wang2002probabilistic,wong2006recurrent,wang2008decision}, long-term memory \citep{hopfield1982neural,chaudhuri2019bipartite,ramsauer2020hopfield}, working memory \citep{durstewitz2000neurocomputational,curtis2003persistent,deco2003attention,barak2014working,seeholzer2019stability}, and the performance of simple cognitive tasks \citep{driscoll2022flexible}. Attractors have also been observed empirically across several experiments investigating decision-making \citep{kurt2008auditory,lin2014sparse,stevens2015fly} and working memory \citep{gnadt1988memory,constantinidis2001coding,curtis2003persistent}.

\subsubsection{Attractors are mutually exclusive: contractive dynamics discretize the state}
\label{sec:discretization}

An important property of attractors is that they are mutually exclusive: each attractor $\mathbf{a}$ is associated with a basin of attraction $B(\mathbf{a})$, which is the region in state-space such that any state $\mathbf{x}$ in $B(\mathbf{a})$ will necessarily converge through the dynamics into $\mathbf{a}$, in the absence of noise or external perturbations. This division into mutually exclusive basins of attraction thus creates a partition of the state space: one can associate to any state $\mathbf{x}$ the attractor $\mathbf{a}$ corresponding to the basin of attraction $B(\mathbf{a})$ in which $\mathbf{x}$ falls.

As a consequence of this mutual exclusivity, any attractor $\mathbf{a}$ has a dual discrete and continuous nature \citep{jaeger1999continuous}: the symbol or composition of symbols $i(\mathbf{a})$ 
that identify $\mathbf{a}$ among all the other possible attractors in the current dynamics is discrete, while a fixed point $\mathbf{a}$ is associated with a real-valued vector (also called embedding~\citep{bengio2000neural,roweis2000nonlinear,morin2005hierarchical} in the deep learning literature) corresponding to the state of the system at that fixed point. 
If the dynamics are not attractive over all dimensions, the same statement can be made for the subspace that is attractive, which means that this discretization effect need not cover every possible dimension and non-discretized dimensions may represent values in a continuous space.

Note that introducing randomness in the dynamics makes it possible to sample one of the attractors that may be reachable from the current state when that noise is taken into consideration. For example, if the state $x$ is close to the boundary between basins of attraction of attractors $A$ and $B$, a small amount of additive noise would suffice to stochastically sample one destination or the other, with probabilities that would vary depending on how far $\mathbf{x}$ is from the boundary and the specific dynamics in its area (for instance, basin depth or slope).

An example of discrete attractor dynamics in the brain can be found in the auditory cortex. \Citet{bathellier2012discrete} studied firing rate patterns in local neural populations in mice and found abrupt shifts between small (1 - 3) numbers of distinct response modes, with response mode identity across multiple local populations providing a discrete code that allowed sound prompts to be identified with 86.2\% accuracy by a linear classifier (compared to 87.3\% using non-discretized activity). 
Abrupt transitions between neural steady states have also been observed in the rat hippocampus and the zebrafish olfactory bulb \citep{niessing2010olfactory,wills2005attractor}.

\subsubsection{Emergent attractors in task-optimized networks}

To demonstrate how attractors naturally emerge as solutions to cognitive tasks, we briefly summarize relevant results from \citet{sussillo2013opening}, where an artificial recurrent neural network (RNN) was trained to solve a simple memory task. An RNN is a network of artificial neurons which can be connected through recurrent feedback loops. Neurons can also form connections to special input and output units, which allow the network to interface with a task. The connection strength between each directed pair of neurons is parameterized using a scalar weight that modulates the degree to which activity in the first neuron drives future activity in the second, and these weights are optimized in order to minimize error on the task. Like the brain, RNNs have recurrent connections between neurons that define a dynamical system optimized to perform some computation, and are therefore useful models for studying emergent neural dynamics.

\begin{figure}[ht]
    \centering
    \includegraphics[width=\linewidth]{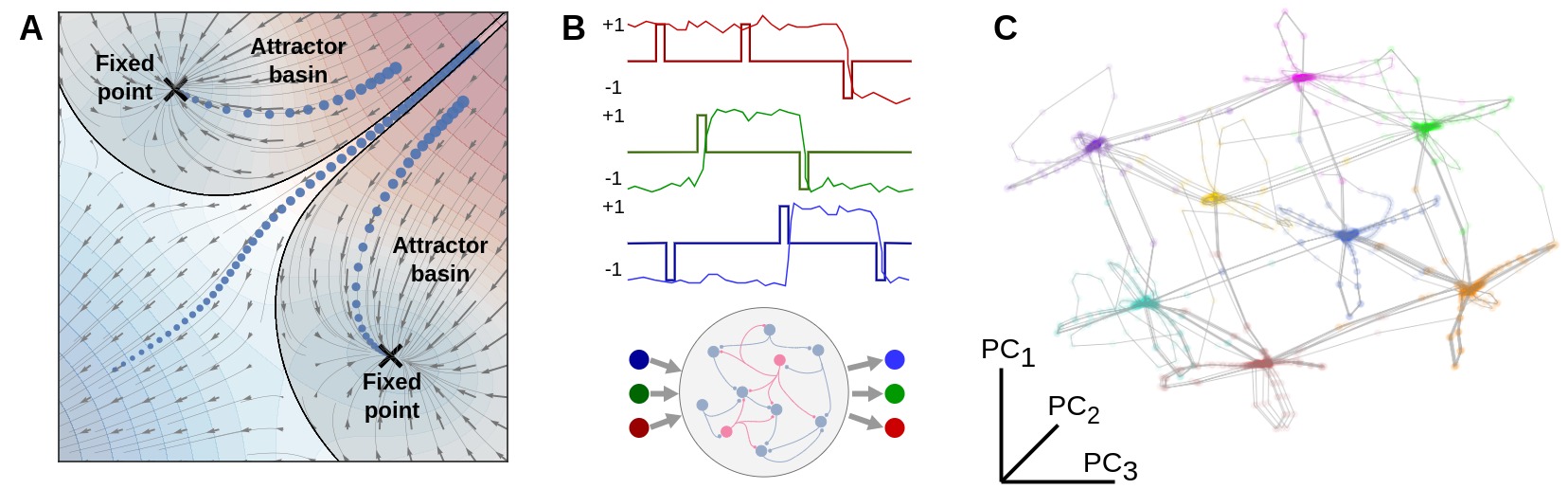}
    \caption{\textbf{Attractor dynamics in neural networks.} \textbf{A.} Attractors in a 2D state space. When a trajectory enters an attractor's basin, it begins to converge to the attractor and remains there until sufficient external input or intrinsic noise allows it to escape. \textbf{B.} \citet{sussillo2013opening} train an artificial recurrent neural network to solve the 3-bit flip-flip memory task. In this task, the model must continuously output the sign of the most recent binary spike on 3 separate input channels. \textbf{C.} Fixed point attractors emerge in the learned dynamics of a recurrent neural network (RNN) as a solution to the task. Each of the attractors corresponds to one of the 8 ($2^3$) possible bit configurations, providing a stable memory state from which the output can be continuously read out. The plot shows a trajectory in the RNN's state-space for changing inputs, where points along the trajectory are colored according to the correct output. The dimensionality of the state space was reduced using Principal Component Analysis (PCA) for visualization.}
    \label{fig:attractors}
\end{figure}

\citet{sussillo2013opening} train an RNN on the 3-bit flip-flop task (\cref{fig:attractors}b), in which the network must learn to continuously output the sign ($+1$ or $-1$) of the last binary spike across 3 input channels (which we can call the ``red'', ``green'', and ``blue'' channels). For instance, following the input sequence [red=+1, green=-1, blue=+1], the correct output should be the vector \{red=+1, green=-1, blue=+1\}. If the next input spike was red=-1, the new output would change to \{red=-1, green=-1, blue=+1\}. Importantly, while each input spike only has a short duration, the network must continuously output the value of each channel's last spike, which imposes a memory demand.

When \citet{sussillo2013opening} inspected the learned dynamics of the RNN, they found that it solved the task through the use of fixed point attractors. Since the number of possible outputs is $2^3 = 8$, the model represented each of these using an attractor. Due to their stability, the model was then able to continuously read out from whichever attractor the trajectory had most recently converged to. Whenever a new spike appeared in one of the input channels (with a value different from that channel's previous spike), the state escaped the current basin of attraction and followed transient dynamics towards the attractor for the new output. This simple task demonstrated how attractor dynamics can naturally emerge in neural networks and implement nontrivial computations, such as those involving transitions between discrete memory states.

\section{An information theoretic dynamical systems perspective on conscious experience}
\label{sec:consious_dynamics}

The main contribution of our paper will be to argue that a dynamical systems model of consciousness with state attractors can account for the communicable, rich, and ineffable aspects of experience that we discussed in \cref{sec:intro}. 
Throughout this section, we will use an information theoretic perspective to characterize richness as information, ineffability as information loss,  communicability as information retention, and we will deploy both the notions of Shannon information and Kolmogorov complexity \citep{kolmogorov1965three}. 
We will illustrate how information loss arises from dimensionality reduction implemented by attractor dynamics.
We show how our model links the problem of accounting for ineffability to the Global Workspace Theory, which predicts that only representations with sufficient amplification and temporal duration (i.e., attractor states) can be broadcast to the rest of the brain for downstream verbal-behavioral reporting.
Finally, we will generalize the notion of ineffability by discussing multiple forms of information loss in intra-personal and inter-personal communication pathways, going beyond the specific case of information loss between working memory and verbal report.

To contextualize our argument, we begin by drawing on existing work to highlight several connections between state attractor models and conscious experience.

\subsection{Motivating attractor dynamics as a model for conscious experience}
\label{sec:motivating_conscious_dynamics}

\subsubsection{Working memory} 

The contents of working memory are typically considered to be the attended contents of short term memory: a 
function of short term representations held in the brain and context from task information or other executive functioning objectives \citep{engle2002working,cowan2008differences}.
A central claim in many leading theories of consciousness is that what we are consciously aware of is the contents of working memory. 
For example, the Global Workspace Theory \citep{baars1993,baars2005} and its neuronal extension \citep{dehaene1998} state that information becomes conscious by gaining entry into a limited workspace that serves as a bottleneck for the distributed activity present across the brain. Pairs of brain regions are largely isolated from each other and arbitrary point-to-point communication is  only possible via the workspace, which itself can both receive and broadcast information globally. The workspace, then, serves as a hub capable of coordinating brain-wide activity for centralized control and decision-making.
It is easy to see the connection between the concepts of a global workspace and working memory 
(attentional selectivity, influence on executive decision-making, availability to verbal and behavioral reporting processes, limited capacity, arbitrary modalities) and there is little distinction between them in the Global Workspace Theory \citep{dehaene2001towards}. Similarly, the notion of ``access consciousness'' introduced in \citet{block1995} can be framed through the lens of a working memory whose contents are globally accessible across the brain.

The link between working memory and attractor dynamics, in turn, is well established.
Empirical studies have demonstrated that attractor dynamics are ubiquitous in the brain, both across species and levels in the brain’s hierarchy \citep{rolls2010attractor,khona2022attractor}.
The attractor model for working memory postulates that working memory emerges from recurrently connected cortical neural networks that allow representations to be maintained in the short term (on the order of seconds) by self-generated positive feedback \citep{durstewitz2000neurocomputational,curtis2003persistent,deco2003attention,barak2014working,seeholzer2019stability}. 
Attractor dynamics can support both \emph{suppression} of inputs, for example in decision making where the brain state flows rapidly towards a discrete attractor and subsequent inputs or perturbations are discounted, as well as \emph{integration} over inputs, where the incremental response to inputs causes reversible flow along continuous attractor manifolds \citep{redish1996coupled,wang2008decision,khona2022attractor}. Neural winner-take-all (WTA) models implement hybrid analog-discrete computation \citep{wang2008decision,wong2006recurrent}. Robustness, discreteness, and temporal integration of information are all traits apparent in working memory \citep{khona2022attractor}.

\subsubsection{Stability and robustness of conscious states}

As a model of conscious processing, discrete attractor dynamics predict that our experience consists of a sequence of relatively stable states that transition swiftly from one to another. Such types of sequential dynamics have been hypothesized to be a key component of conscious thought and perception \citep{james1892,varela1999,rabinovich2008,tsuda2015}.
Empirically, one of the characteristics that distinguishes conscious vs. unconscious neural representations in psychophysics tasks is that they are significantly more stable in the ``aware'' condition \citep{schurger2015}.
 
Qualitatively, subjects commonly report on the emergence of stable discrete ``choices'' within conscious perception.
For instance, when looking at the Necker cube, subjects only perceive one single interpretation of its structure and orientation rather than a mixture of both possibilities. Occasionally, this interpretation will change to the alternative one, but the change will happen rapidly as an abrupt transition. Similarly, in the case of binocular rivalry, only a single image presented to one of the eyes will be consciously perceived rather than a mixture of the two, and which image is consciously perceived will abruptly change at random times. 
Such cases are characterizable by attractor dynamics that converge to one attractor and remain stable until sufficient input change or noise result in a rapid transition to another attractor.

Input change or noise may also result in basin transitions that occur without complete convergence to attractors. This is familiar in the cases of thought and speech. One common example is thought-disruptive external stimuli, in which external stimuli distract or interrupt one's chain of thought. A less well-known but equally important example is the role of internal time-saving mechanisms. These are active in cases where one does not need to spell something out in full detail. 
For example, in speech production, phonemes are often not fully articulated: this may be understood by noting that once one has arrived at an attractor basin it is disambiguated which point one converges toward \citep{Roessig-intonation}. A similar mechanism may explain the utility of verbal or symbolic thought, where the key may serve as synecdoche for the value.

\citet{schurger2010} suggested that conscious states were associated with increased robustness to noise in psychophysics experiments. A signature of neural representations in the ``conscious'' condition was that they were highly reproducible; given the same stimulus presentation across different trials, patterns of neural activity were similar, so long as the subjects reported awareness of the stimulus. In contrast, patterns of activity during the ``nonconscious'' condition in which subjects were unaware of the stimulus exhibited greater variability. Both robustness to noise and reproducibility of states, in turn, are core properties accommodated by attractor dynamics.

\subsection{Richness and ineffability}
\label{sec:richness_and_ineffability}


\begin{mathbox}[label=math:notation]{Notation}
Let lower case $x$ denote an instance of random variable $X$, $\mathcal{X}$ denote the set of possible states for $X$ with probability distribution $P(X)$, $\sum_{x \in \mathcal{X}} P(X = x) = 1$, $p(x)$ denote $P(X = x)$, expectation $\mathbb{E}_{p(x)} [f(x)]$ denote $\sum_{x \in \mathcal{X}} p(x) f(x)$, and likewise for other variables. We restrict function domains to discrete variables including floating point representation of reals. $[n]$ denotes the list of natural numbers $1, \dots, n$. 
\end{mathbox}

What is meant by the richness of experience? 
Intuitively, whilst we find it easy to communicate certain aspects of our mental state, we struggle to convey their full content or meaning. 
One can consider color as an example. 
We are tempted to think of color space as a simple 3-dimensional surface, on the basis of perceptual similarity judgments that people tend to make. However, there is a far richer and higher dimensional structure to experiencing color. 
For instance, most people would describe the color “red” as warm and aggressive. There are myriad associations that we make with various colors that are not functions of their nominal definitions, and all of these associations as a whole contribute to the richness of the experience \citep{chalmers2010character}.

Broadly, richness means having a lot; the condition of being ``well supplied or endowed'' \citep{def}. In the context of mental state attribution, richness gauges the amount of specificity---detail, texture, nuance or informational content---contained by a mental state. It is a common principle in aesthetics that experience is rich (a picture speaks a thousand words), and many philosophers acknowledge that conscious states at least appear to be highly detailed, nuanced, and contentful \citep{Tye2006-TYENCR,Chuard2007-CHUTRO-2, block1995}, though some take this appearance to be ultimately illusory \citep{dennett1993consciousness, Cohen2016-COHWIT-3}. 

This conception of richness corresponds well to the mathematical notion  formalized by Shannon \citep{shannon1948mathematical}, where richness of a random variable $X$ is given by its entropy $H(X)$.
Here, a random variable represents a state type, e.g., experience of some face or other. To say that such a variable is high in entropy is to say that the number of values it could take (the number of possible states the system could be in, e.g., the different experiences of faces one could possibly have) is relatively large and the probability distribution over these is relatively flat,
and thus the state is unpredictable. Specifically, Shannon entropy $H(X)$ quantifies the average number of bits (answers to yes-or-no questions) required to specify which state $X$ takes as a measure of informational content.

The notion of ineffability is closely related. In popular usage, ineffable can be defined as ``too great for words'' \citep{def2}. The concept is often used in theological contexts, but it has been applied to descriptions of qualitative experience since at least \citet{dennett1993consciousness}. Given the term's theological associations, the claim that experience is ineffable might sound like a profession of dualism: consciousness is something magic that no physicalist theory can account for. However, strictly speaking, to claim that experience is ineffable is simply to claim that its informational content exceeds what we can remember or report. Much hinges on what exactly we mean by ``can remember or report''. Of course, one can say a thousand words, so the fact that a picture speaks that many words does not necessarily make a picture ineffable. 
Below, we will develop tools to allow us to precisely refine the senses of ineffability at issue, and we will see that experience is ineffable in multiple senses (though none of them need involve magic or anything anathema to physicalist theories).

We propose that ineffability corresponds to the mathematical notion of information loss when trying to express a conscious state in words. Given a function that processes an input variable $X$ and produces an output variable $Y$, information loss of the input incurred by the output is measurable by conditional entropy $H(X | Y)$, or entropy of the input variable given the output variable. Intuitively, conditional entropy $H(X | Y)$ measures how well $Y$ describes $X$: how much uncertainty remains about the value of $X$, once the value of $Y$ is given. Conditional entropy $H(X | Y)$ is mathematically equivalent to the  entropy of the input $X$ minus the mutual information between input and output, $H(X | Y) = H(X) - I(X ; Y)$, where the latter is a measure of information shared between them; the amount of information about the state of one variable obtained by observing the state of the other. 
Note the difference between conditional entropy and mutual information: mutual information is how much uncertainty one random variable removes from another, while conditional entropy describes how much uncertainty remains in the first variable after the value of the second is given. Later we will also make use of joint entropy, $H(X,Y)$, which is the amount of information needed on average to specify the states of both $X$ and $Y$.

Usefully, quantifying inefability in this manner allows us to offer a precise definition of effability as the negation of ineffability. Where ineffability is given by $H(X | Y)$, negating ineffability gives effability: $- H(X | Y) = I(X ; Y) - H(X)$. Recalling that entropy is a measure of uncertainty or spread in a probability distribution, the smaller $H(X | Y)$ is, the less uncertain $X$ is given $Y$, the less information is lost, and the more effable or communicable $X$ is via $Y$. We may say that given a variable $X$ with entropy $H(X)$, its effability to variable $Y$ scales with the amount of shared information $I(X ; Y)$.
Finally, since entropy $H(X)$ is recoverable as $H(X) = H(X | C)$ for any constant variable $C$, richness may be considered the special case of ineffability where the output state is a constant. 

In the foregoing we draw on the framework of Shannon information, but there are advantages, for our purposes, to using Kolmogorov information \citep{kolmogorov1965three} as an alternative way to characterize richness and ineffability. 
In the Kolmogorov formalism, richness of a state $x$ corresponds to its complexity $K(x)$, which is the length in bits of the shortest program written in a general programming language that outputs $x$ and halts. Ineffability then corresponds to conditional Kolmogorov complexity of an input $x$ given an output $y$, $K(x | y)$, the length of the shortest program needed to produce $x$ if $y$ is given, or intuitively the complexity of $x$ minus the number of bits that can be saved from knowing $y$, which is the Kolmogorov analog of Shannon information loss as conditional entropy or entropy minus mutual information. 
Note that since Kolmogorov complexity is defined on strings of bits, we restrict the domain of our functions to discrete variables and assume that floating point representation is used to encode real values (Box \ref{math:notation}), which is natural for variables modelled or stored with computer memory. Construction of the computational model thus incurs a separate form of information loss resulting from discretization of real-valued continuous-time observations of neural state. 

\begin{figure}
    \centering
    \includegraphics[width=0.75\textwidth]{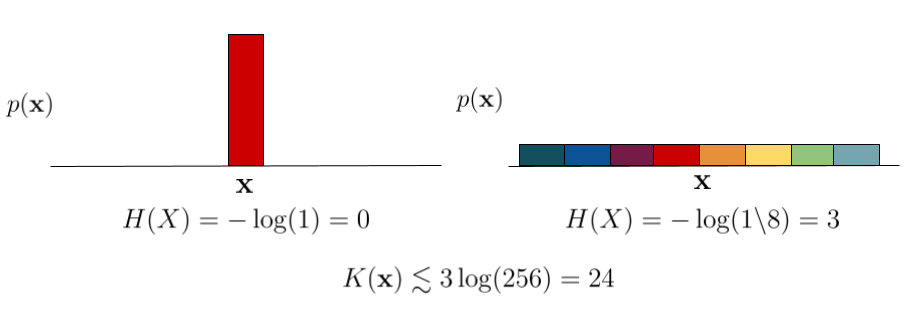}
    \caption{Illustrating Shannon entropy and Kolmogorov complexity for discrete color distributions.
    Entropy (\cref{eq:entropy}) involves an expectation over the states of stochastic variable $X$ whereas Kolmogorov complexity (\cref{eq:k_complex}) is defined for an instance of state, $x$. 
    The distribution on the left has non-zero mass in one state and is the minimum entropy distribution; the distribution on the right is uniform over states and is the maximum entropy distribution for 8 states. Assume a universal RGB representation for colors where each RGB component ranges between 1 and 256. Without assumptions on the distribution over colors, the Kolmogorov complexity of each state is no greater than 24 (excluding program overheads) since color can be represented with 3 8-bit binary sequences, 
    but may be lowered for smaller RGB values that do not require 8 bits if an optimized number encoding scheme is used \citep{grunwald2003kolmogorov}.
    Whereas entropy is the same for the same probability distribution over \emph{any} states, Kolmogorov complexity would increase for states whose values are algorithmically more difficult to construct.
    }
    \label{fig:colour}
\end{figure}

Shannon entropy and Kolmogorov complexity are closely related metrics of richness, and are described in more detail in Box \ref{math:defs} and \cref{fig:colour}. If the probability distribution over states is given, taking an expectation over the distribution on Kolmogorov complexity of its states allows Shannon entropy to be approximately recovered \citep{grunwald2004shannon}. 
Under either framework, richness is characterizable as information measured in bits, ineffability as information loss or richness reduction, and communicability and ineffability are neither separate nor boolean traits, but direct opposites of each other and varying on a scale. 

A major difference between Shannon entropy and Kolmogorov complexity is the former is defined given a probability distribution over variable states, 
whereas the latter is defined on individual states without assuming a given probability distribution. 
Shannon entropy is defined for a random variable given its probability distribution, measuring the average information carried by states, while Kolmogorov complexity is a measure of the information carried by an individual state $x$, without dependency on the probability distribution of a random variable ranging over it.
Knowledge of the distribution over a variable's states is generally a non-trivial assumption.
The distribution may be undefined or highly privileged information in itself (that is, the meta-distribution over the distribution's parameters is rich). 
Consider, for example, measuring the amount of information in a book by considering the set of all possible books and the distribution over them \citep{grunwald2003kolmogorov} or the information in a temporal snapshot of a high-dimensional brain state by considering the distribution over all possible states. In these cases, we want a way to measure informational content that does not require knowledge of a hard-to-specify distribution.
This is especially salient for us where inter-personal ineffability is concerned. Even if we assume that a brain's parameters fully determine the distribution over its own states (and so in some sense individuals have direct access to their own distributions), still individuals cannot have this level of knowledge of the distributions of their interlocutors' brains.  Explicitly allowing the communicator's distributional parameters to be unknown is therefore convenient for characterizing inter-personal ineffability from the perspective of the listener.

A second drawback of Shannon's framework is that entropy is a measure of statistical determinability of states as opposed to difference in absolute states; information is fully determined by the probability distribution on states and unrelated to the meaning, structure or content of individual states \citep{grunwald2003kolmogorov}. 
For example, consider again a case where we want to measure inter-personal ineffability, as a relationship between a communicator's experience and a listener's. Conditional entropy of the communicator's experience given the listener's experience is low if the pairing is statistically unique, regardless of the semantic correspondence between experiences, whereas conditional Kolmogorov complexity is concerned with the difficulty of reconstructing the communicator's experience given the listener's experience, i.e. absolute difference, which corresponds more closely to the lay definition of ineffability. For example, it might just happen to turn out that whenever Alice thinks and talks about tennis, Bob almost always thinks about Beethoven. In this case, conditional entropy will be low, but conditional Kolmogorov complexity will be high, and therefore suited to capture the absolute difference between their experiences.
For these reasons, we argue that particularly in the case of inter-personal communication, Kolmogorov complexity should be used to characterize richness and ineffability of experiences. However, Shannon entropy is functionally equivalent if the distribution is given, and we will refer to both frameworks. 


\begin{mathbox}[label=math:defs]{Metrics for richness and ineffability}


Shannon entropy is given by
\begin{align}
H(X) = \mathop{\mathbb{E}}_{p(x)} [- \log p(x)]. \label{eq:entropy}
\end{align}
If variable $Y$ is produced by processing $X$, $y = f(x)$, with joint distribution denoted by $p(x, y)$ and $f$ stochastic in the general case, then information loss from $X$ to $Y$ is given by conditional entropy $H(X | Y) = H(X) - I(X; Y)$, where $I(X; Y)$ denotes Shannon mutual information between variables,
\begin{align}
I(X; Y) = \mathop{\mathbb{E}}_{p(x, y)} \bigg[ \log \frac{p(x, y)}{p(x) p(y)} \bigg],
\end{align}
and $H(X | Y)$ is given by 
\begin{align}
H(X | Y) = \mathop{\mathbb{E}}_{p(x, y)} [ -\log p(x | y) ].
\end{align}

The Kolmogorov complexity of a state $x$, $K(x)$, is the length $l(z)$ in bits of the shortest binary program $z$ that prints $x$ and halts.
Specifically, let $U$ be a reference prefix universal machine. The prefix Kolmogorov complexity of $x$ is
\begin{align}
K(x) = \min_z \{l(z) : U(z) = x, z \in \{0, 1\}^*\} \label{eq:k_complex}
\end{align}
Conditional Kolmogorov complexity is the length of the shortest program that takes $y$ as an input, prints $x$ and halts. It is given by
\begin{align}
K(x) - I(x : y) = K(x | y^*) \stackrel{+}{=} K(x | y, K(y)) \stackrel{\log}{=} K(x | y) \label{e:k_ineff}
\end{align}
where $I(x : y)$ denotes Kolmogorov mutual information between states, $y^*$ denotes the shortest program that produces $y$ and halts, standard notation $\stackrel{+}{=}$ and $\stackrel{\log}{=}$ are used to denote equality up to constant and logarithmic factors respectively
\citep{grunwald2004shannon,li2008introduction}.
As \cref{e:k_ineff} shows, $K(x | y^*)$ and $K(x | y)$ are comparable and either may be used to characterize information loss; in subsequent sections we will generally refer to $K(x | y)$. \\

Kolmogorov complexity has several intuitive properties and similarities with Shannon entropy. Conditioning on more data cannot increase the information of a state, $K(x | y) \leq K(x)$, as $y$ is utilized if it allows for a shorter program and otherwise ignored. 
If $y$ merely copies $x$ there is no information loss, $K(x | y) \stackrel{+}{=} 0$.
Under the Shannon framework, $H(X | Y) = 0$ if $X = Y$ but also more generally if each state $y$ corresponds to a unique $x$.
\Cref{fig:colour} illustrates Shannon entropy and Kolmogorov complexity for a toy example. Shannon entropy is concerned with statistical determinability of a random variable given knowledge of its probability distribution, whereas Kolmogorov complexity can be considered as a more tabula rasa (not knowing the distribution) measure of richness of a particular value of this variable.
Shannon entropy and Kolmogorov complexity are related by the following constraints \citep{grunwald2004shannon}:
\begin{align}
0 &\leq (\mathbb{E}_{p(x)} [K(x)]) - H(X) \stackrel{+}{\leq} K(p), \label{eq:k_ent}\\
I(X; Y)  &\stackrel{+}{=} \mathbb{E}_{p(x, y)} [I(x : y | p)], \\
I(X; Y) - K(p) &\stackrel{+}{<} \mathbb{E}_{p(x, y)} [I(x : y) ] \stackrel{+}{<} I(X; Y) + 2K(p),
\end{align}
which conveys how Kolmogorov complexity pays a penalty for not assuming knowledge of the distribution, since it must be encoded within the program.

\end{mathbox}


\subsection{Intra-personal ineffability}
\label{sec:intra}

\begin{figure}[ht]
    \centering
    \includegraphics[width=\linewidth]{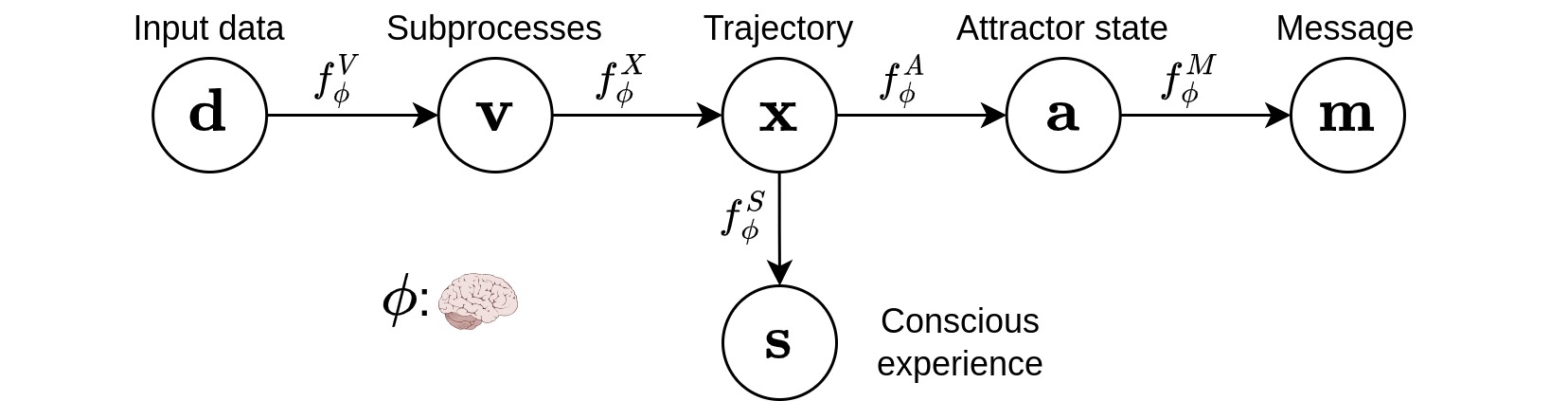}
    \caption{\textbf{A model of intra-personal ineffability.} Information is channelled through the stages of input ($\mathbf{d}$), subprocesses state ($\mathbf{v}$), working memory ($\mathbf{x}$, $\mathbf{a}$), conscious experience ($\mathbf{s}$) and verbal report ($\mathbf{m}$). 
    A trajectory $\mathbf{x}$ in the state-space of working memory follows attractor dynamics, converging near an attractor $\mathbf{a}$. Each step transforming one variable to another is executed by the dynamics of the individual's brain, which is determined by parameters $\phi$. Conscious experience $\mathbf{s}$ is a function of the subject's cognitive parameters $\phi$ and working memory trajectories $\mathbf{x}$, and encodes the experience's meaning. 
    }
    \label{fig:intra-personal}
\end{figure}

In this section we will develop our model of intra-personal ineffability, that is, ineffability between stages of processing within a single experiencer. We will be concerned with the following variables: Let $X$  (with value $\mathbf{x}$) be a trajectory of neural activities that determine working memory content and conscious experience, and let it consist of a sequence of transient states $X_t$ for $t \in [T]$, where length $T$ is fixed and sufficiently large such that all trajectories terminate near an attractor state. 
Let $A$ (with value $\mathbf{a}$) denote the terminating attractor, $S$ (with value $\mathbf{s}$) denote the conscious experience, $D$ (with value $\mathbf{d}$) denote external input datum, let $V$ (with value $\mathbf{v}$) denote a list of $N$ subprocess states $V_n$ for $n \in [N]$ and fixed $N$ that comprise computation affecting working memory trajectory $X$, and let $M$ (with value $\mathbf{m}$) denote the verbal report or output message of the individual. 
In addition, let $\phi$ denote the brain's synaptic weights that parametrize its dynamics.
These variables are connected by a computation graph of functions (\cref{fig:intra-personal}), given by $\mathbf{v} = f^{V}_\phi(\mathbf{d})$, $\mathbf{x} = f^{X}_\phi(\mathbf{v})$, $\mathbf{a} = f^A_\phi(\mathbf{x})$, $\mathbf{s} = f^S_\phi(\mathbf{x})$ and $\mathbf{m} = f^M_\phi(\mathbf{a})$. The functions $f^A_\phi$ (returns final attractor state) and $f^S_\phi$ (outputs conscious experience that is fully determined by $\mathbf{x}$) are deterministic while  $f^M_\phi$, $f^{V}_\phi$ and $f^{X}_\phi$ are generally stochastic, meaning outputs may be dependent on hidden stochastic variables within the function that encode historical states or neural processing noise.
Not speaking is encoded by a state of $V$ corresponding to ``no verbal report''. 
Subscripting with $\phi$ denotes that function behavior is determined by cognitive parameters $\phi$.
The computation graph defines a joint probability $p_\phi(\mathbf{d}, \mathbf{x}, \mathbf{s}, \mathbf{a}, \mathbf{m})$, from which conditional and marginal probability distributions on individual variables may be obtained.
Entropy $H_\phi$ is also parameterized since it depends on $p_\phi$.
Finally, denote the transient state by $\bar{X}$, where $p_\phi(\bar{\mathbf{x}}) = \frac{1}{T} \sum_{t \in [T]} P_\phi(X_t = \bar{\mathbf{x}})$ is the probability that any transient state takes the value $\bar{\mathbf{x}}$.

Our dynamical systems model of working memory distinguishes between two kinds of working memory state, attractor states and transient states, where the latter includes all time-varying states occupied by the system and the former corresponds to system output, or the accessible contents of working memory~\citep{khona2022attractor}.
Our model remains neutral about whether conscious states correspond to working memory trajectories, transient states or attractor states but allows conscious state to be more generally a deterministic function of these states, thus conveying part of their information. 
Specifically, since $\mathbf{s} = f^S_\phi(\mathbf{x})$, conscious experience is not restricted to be identical to transient working memory states or attractor states, but is the output of a deterministic function of the trajectory through working memory states, where the function depends on cognitive parameters $\phi$. While we will not focus on the implementation details of how conscious experiences might relate to neural processes, intuitively $\mathbf{s}$ can be thought of as a vector of real numbers representing one point in an abstract space of possible experiences.
Subsequently, information theory gives us the ability to reason about the relative richness and ineffability of conscious experience based on the computation graph, without needing implementation details of the functions.

\subsubsection{Information loss from attractor dynamics}
\label{sec:attractor_info_loss}



The relation of trajectories $\mathbf{x}$ to a smaller subset $\mathbf{a}$ of attractor states is a defining characteristic of attractor dynamics, whether the subset consists of a discrete number of fixed points or a set of states that trace out a complex shape such as a curved manifold. In this section, we argue that the presence of attractor dynamics decreases the richness of working memory states and conscious experience. We will identify two related effects. First, at the level of comparison between systems, the presence of attractors concentrates the probability mass of transient states onto a smaller subspace, reducing the richness of transient states.
Second, we show that at the level of comparison between states, since attractor states are less rich than transient states in general and the former constitute outputs of the system, the richness of attractor states limits the richness of downstream variables.

Since dynamics are characterized by the flow of transient states towards an attractor in $\mathcal{A}$ followed by persistent membership in $\mathcal{A}$, and attractors $\mathcal{A}$ typically constitute a significantly smaller subset of all possible transient states $\mathcal{\bar{X}}$ \citep{khona2022attractor}, 
the presence of attractors decreases the richness of transient states $H_\phi(\bar{X})$.
Since entropy is a measure of distributional spread, dynamics with larger non-attractor transient state spaces $\mathcal{\bar{X}} \setminus \mathcal{A}$, implying more time spent in non-attractor states, 
yield richer distributions over transient states $P_\phi(\bar{X})$; conversely, faster convergence to attractors and more time spent at attractors yields lower $H_\phi(\bar{X})$.
In turn, reducing the richness of transient states limits the richness of full trajectories and conscious experience (Box \ref{math:trans_rich}). 

\begin{mathbox}[label=math:trans_rich]{Implications of reducing transient state richness}
Reducing the richness of transient states $H_\phi(\bar{X})$ also reduces a ceiling on the richness of full trajectories $H_\phi(X)$, since $H_\phi(X) = H_\phi(X_1 \dots X_T) \leq $ \linebreak $\sum_{t \in [T]} H_\phi(X_t) \leq  T (H_\phi(\bar{X}) + C)$ by the addition rule of entropy, where constant $C = \max_{t \in [T]} (H_\phi(X_t) - H_\phi(\bar{X}))$ limits the maximum deviation of entropy between individual timesteps and the temporal average.
This in turn reduces a ceiling on the richness of conscious experience as $H_\phi(S) \leq H_\phi(X)$. The latter can be shown as follows: the joint entropy $H_\phi(S, X) =  H_\phi(X) + H_\phi(S | X) = H_\phi(X)$ since $f_\phi^S$ is deterministic, i.e., $H_\phi(S | X) = 0$. $H_\phi(X) = H_\phi(S, X) = H_\phi(S) + H_\phi(X | S)$ and Shannon entropy is non-negative, thus $H_\phi(S) \leq H_\phi(X)$.
\end{mathbox}

This might seem to be an artefact of the Shannon approach, which directly concerns features of the distribution. 
However, the same reasoning applies under Kolmogorov's formalism if the probability distribution is known, because loosely speaking, knowing the distribution gives the encoder a short-cut: 
expected Kolmogorov complexity $\mathbb{E}_{p_\phi(\bar{\mathbf{x}})} K(\bar{\mathbf{x}} | p_\phi)$ is equivalent to entropy $H_\phi(\bar{X})$ up to an additive constant if the distribution is given  (\cref{eq:k_ent}). 
Intuitively, this is because the shortest lossless descriptor of $\bar{\mathbf{x}}$, given knowledge of the distribution $P_\phi(\bar{X})$ and thus support $\mathcal{\bar{X}}$, has length $- \log p(\bar{\mathbf{x}})$ under Shannon's noiseless coding theorem \citep{grunwald2004shannon}. Given knowledge of $P_\phi(\bar{X})$, $- \log p(\bar{\mathbf{x}})$ bits are all that is additionally needed to determine the state using a descriptionally simple (but not necessarily computationally short) computer program. 

Thus far we have described how the presence of attractors can decrease the richness of transient states overall, i.e., as a matter of comparing between systems (e.g., two brains). We turn now to a second way in which attractors reduce richness, as a matter of comparison between states in a given system.

Global Workspace Theory postulates that the access of representations from working memory by diverse processes across the brain depends on the representations being \emph{amplified and maintained over a sufficient duration}, for instance for a minimum of approximately 100ms \citep{dehaene2001towards}.
In the language of the attractor framework, this amounts to the claim that the variable released to downstream processes such as verbal-behavioral reporting and long-term memory is $A$, not $X$. Crucially, attractor states are strictly less rich than trajectory states $H_\phi(A) < H_\phi(X)$, as explained in Box \ref{math:gwt}.
Thus selective release of attractor working memory states to downstream processing functions such as $f_\phi^M$ implements an information bottleneck that limits the richness of downstream inputs. 
This constitutes an important source of ineffability, where our in-the-moment experiences $S$ are
richer than our later recollections, since richness of experience is upper bounded by the richness of trajectories (i.e. $H_\phi(X) \geq H_\phi(S)$, Box \ref{math:trans_rich}), so the higher the richness of trajectories, the higher the ceiling on information loss from conscious experience to the attractor state and downstream variables. This will be relevant to our discussion of phenomenal overflow \citep{block2007overflow} below.
In practice one would expect the magnitude of information loss from trajectory $X$ to working memory output $A$ to be significantly large, since trajectories are sequences of brain states specifying the activity of billions of neurons, whereas working memory appears to be limited to representing a handful of items \citep{sperling1960information}, which gives us a clue to the magnitude of the bottleneck.



\begin{mathbox}[label=math:gwt]{Richness of attractors strictly less than richness of trajectories}
As the full trajectory determines the attractor it terminates in, $f^A_\phi$ is a deterministic function. It follows that $H_\phi(A | X) = 0$. We also know that $H_\phi(X | A) > 0$ since multiple possible trajectories terminate in the same attractor state. Our result follows from this asymmetry. By the general relationship between joint and conditional entropy we have $H_\phi(X, A) = H_\phi(X) + H_\phi(A | X)$.  Since $H_\phi(A | X) = 0$ we have $H_\phi(X, A) =  H_\phi(X)$. Re-applying the  relation between joint and conditional probability we also have $H_\phi(X, A) = H_\phi(A) + H_\phi(X | A)$. From these observations together we know that $H_\phi(A) + H_\phi(X|A) = H_\phi(X)$. Since $H_\phi(X | A) > 0$, this yields  $H_\phi(A) < H_\phi(X)$. 
\end{mathbox}

\subsubsection{Information loss at verbal report}
\label{sec:message_info_loss}

The ineffability of an experience is perhaps most obvious when we attempt to put it into words, due to the highly compressed nature of language \citep{kirby2015compression}. 
From the computation graph, we can say that ineffability or information loss from conscious experience to verbal report is at least as great as information loss from conscious experience to working memory attractor (Box \ref{math:verbal_report}).
Additionally, it would be reasonable to assume information losses $H_\phi(A | M)$ and $H_\phi(S | M)$ are strictly positive (i.e., $ H_\phi(A | M) > 0$ and $H_\phi(S | M) > 0$) if message $M$ is a low-dimensional symbolic variable (such as a few words) whereas $A$ and $S$ are snapshots of working memory and conscious experience, since conditional entropy is strictly positive if every mapping is either one-to-one or one-to-many and there is at least one case of the latter.
While it might appear that language is rich, note that $n$ characters with an alphabet of 256 possible characters require no more than $8n$ bits to represent, whereas neural state is determined by the activity of up to approximately 100 billion neurons \citep{herculano2009human}. 

Information loss from attractor $A$ or conscious experience $S$ to verbal message $M$ means the latter do not fully identify the former, and instead divide the space of attractors and conscious experiences more coarsely. 
For instance, saying that one ``saw a fat cat'' leaves out significant details about the specific attractor that generated the message, which would be difficult to communicate fully (e.g., the cat's color, size, pose, the surrounding environment, etc.). 
Positive information loss $H_\phi(S | M)$ implies it is generally impossible to  recover the conscious experience from the verbal message with certainty. Note that as long as $H_\phi(A | M)$ is strictly positive, this means that conscious experience is somewhat ineffable to verbal report even if we identify conscious experience with working memory attractor states.

\begin{mathbox}[label=math:verbal_report]{Ineffability of conscious experience to verbal report}
From the computation graph, $S - X - A - M$ form a Markov chain ($S$ is conditionally independent of $A$ if given $X$), thus $S - A - M$ is also a Markov chain ($S$ is conditionally independent of $M$ if given $A$). Thus $I_\phi(S; A) \geq I_\phi(S; M)$ from the data processing inequality theorem, implying $H_\phi(S) - H_\phi(S | A) \geq H_\phi(S) - H_\phi(S | M)$ and $H_\phi(S | M) \geq H_\phi(S | A)$.
\end{mathbox}




An additional source of ineffability is that attractors can have more complex and high-dimensional structure than simple fixed points, which is common in high-dimensional systems. Such a system would exhibit increased richness of attractor state $H_\phi(A)$ and increased ineffability, as the same richness of messages $H_\phi(M)$ and an increase in joint entropy $H_\phi(A, M)$ implies an increase in information loss $H_\phi(A | M)
$, since $H_\phi(A, M) = H_\phi(M) + H_\phi(A | M)$.



\subsubsection{Hierarchical attractor dynamics}

The brain is hierarchical in nature with many levels of spatial and temporal organization that can be studied, ranging from molecular and synaptic activity to local networks and large-scale networks \citep{changeux1989neuronal}.
Attractor dynamics appear to be ubiquitous across organizational levels and cortical regions of the brain, with processing in the neocortex hypothesized to support many attractor networks each concerned with a different type of processing (executive function and working memory, general short-term memory, long-term memory etc.) \citep{rolls2007memory,rolls2010attractor, khona2022attractor}.
The presence of multiple weakly coupled neocortical attractor networks yields benefits including specialization and increased memory capacity, and in addition has ramifications for understanding conscious experience. 

Anatomically, the inferior temporal cortex is an example of a sensory processing area that responds discriminatively 
to novel stimuli,
whereas the prefrontal cortex is implicated in maintaining attention-modulated projections of such representations in working memory \citep{rolls2007memory,miller1993activity,renart1999recurrent}. 
Neural activity in both regions maintains persistence over time and exhibits attractor dynamics, but the content of sensory memory is akin to the state of a worker subprocess whereas the content of working memory corresponds to the state of executive control; working memory representations exhibit increased temporal stability, persisting for longer durations of up to several seconds, and provide top-down feedback to diverse regions of the brain, including the inferior temporal cortex \citep{rolls2010attractor,chelazzi1999serial,bushnell1981behavioral}.
The ability of the prefrontal attractor to 
stabilize in its high firing rate attractor state is attributable to positive feedback from strong internal recurrent connections that suppress incoming stimuli \citep{renart1999recurrent}.
The need to maintain information in working memory during periods where new stimuli may be perceived exemplifies why working memory and subprocess memory necessitate distinct attractor networks \citep{rolls2010attractor,rolls2007memory}.

The well-known Sperling experiments \citep{sperling1960information} illustrate different dynamics in working memory and sensory memory processes, notably in terms of duration (a few seconds or less after the brief visual presentation of an array of letters, only letters that have been consciously attended to remain reportable) and capacity (sensory memory is capable of holding rich information pertaining to all digits whereas the number of reportable items was limited to approximately 4).
Numerous studies have demonstrated the short-lived nature of representations in sensory memory and the importance of top-down feedback, as backprojected attention appears necessary to avoid exponential decay in sensory memory representations \citep{rolls2010attractor,cohen1998competition,tiitinen1994attentive,rolls1994processing}.

The limits imposed on the richness of working memory state by 
subprocess memory states may be illustrated in an information theoretic manner by considering that the latter is an input to the former (Box \ref{math:subprocess}).

\begin{mathbox}[label=math:subprocess]{Richness of subprocess states constrains richness of conscious experience}
Extracting the stochasticity in $f_\phi^X$ into an input variable $\omega$, meaning assuming that computation of $X$ is cast as $X = \hat{f}_\phi^X(V, \omega)$ where $\hat{f}_\phi^X$ is deterministic, the richness of $X$ is bounded as $H_\phi(X) \leq H_\phi(V_1, \dots, V_N, \omega) \leq \sum_{n \in [N]} H_\phi(V_n) + H_\phi(\omega)$ due to deterministic data processing and addition rule of entropy. That is, given a limit on the richness of noise $H_\phi(\omega)$, a ceiling on the richness of working memory trajectories $H_\phi(X)$ scales with the richness of the subprocess states that constitute its inputs. In turn this restricts ceilings on the richness of downstream variables such as conscious experience and working memory attractors (Box \ref{math:trans_rich}). 
\end{mathbox}











\subsection{Inter-personal ineffability}
\label{sec:inter}

\begin{figure}[ht]
    \centering
    \includegraphics[width=\linewidth]{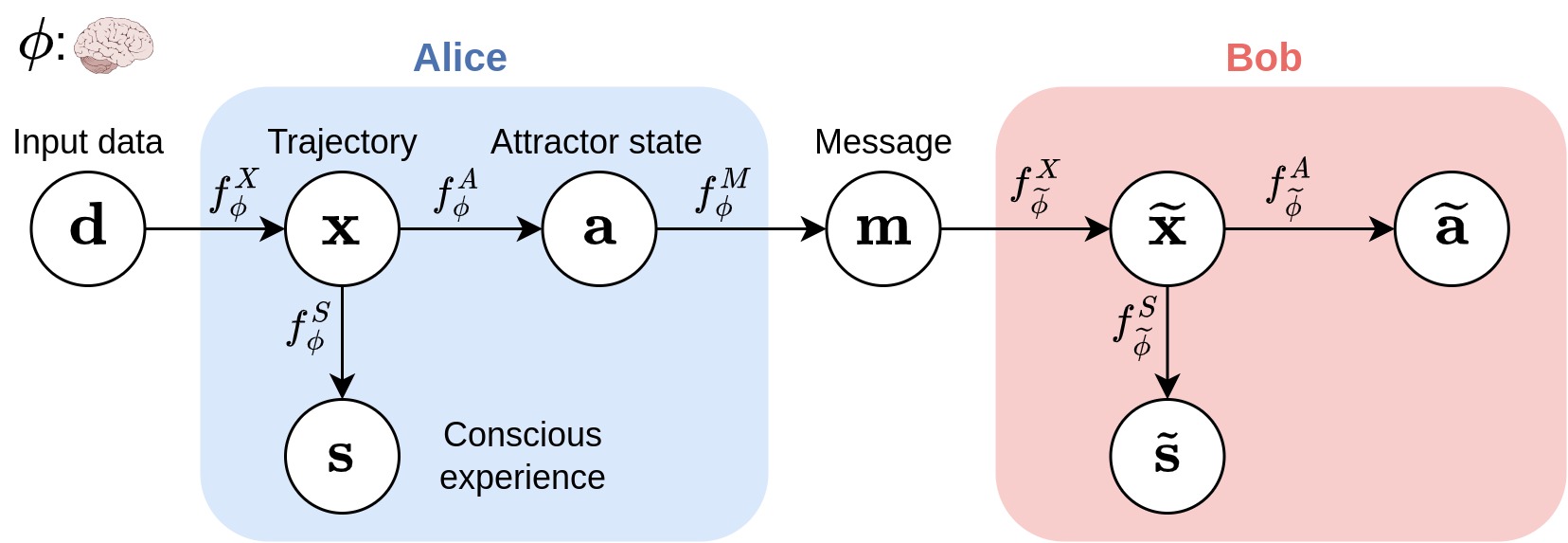}
    \caption{\textbf{A model of inter-personal ineffability.} We model the communication pipeline between a speaker Alice and a listener Bob. A trajectory $\mathbf{x}$ in Alice's state-space of working memory follows attractor dynamics, converging near an attractor $\mathbf{a}$. Alice then attempts to communicate the experience with a message $\mathbf{m}$. On Bob's end, the message is decoded and influences his working memory trajectory $\tilde{\mathbf{x}}$, which in turn converges near an attractor $\tilde{\mathbf{a}}$. Each step transforming one variable to another is executed by the dynamics of the subject's brain, denoted by $\phi$ for Alice and $\tilde{\phi}$ for Bob. Conscious experiences $\mathbf{s}$ and $\tilde{\mathbf{s}}$ are functions of the subject's cognitive parameters $\phi$ and $\tilde{\phi}$ and working memory trajectories $x$ and $\tilde{x}$ respectively, and encode the experience's meaning. We are interested in the ineffability $K(\mathbf{s} | \tilde{\mathbf{s}}, p_{\tilde{\phi}})$ of Alice's conscious experience $\mathbf{s}$ given the experience $\tilde{\mathbf{s}}$  elicited in Bob.
    }
    \label{fig:inter-personal}
\end{figure}

Communication channels are not limited to personal sensory processes and verbal or behavioral reporting processes but extend to channels between individuals. In this section, we will consider communication between two individuals using the model summarized in \cref{fig:inter-personal} in which a speaker, Alice, wishes to communicate her experience to a listener, Bob. We use the same variables as as in prior sections, but denote Bob's variables using a ``$\sim$'' (e.g., $\tilde{\mathbf{s}}$ denotes Bob's conscious experience). Again we assume a computational chain of states $\mathbf{x} \rightarrow \mathbf{a} \rightarrow \mathbf{m} \rightarrow \tilde{\mathbf{x}} \rightarrow \tilde{\mathbf{a}}$ that elicit an experience $\tilde{\mathbf{s}} = f_{\tilde{\phi}}(\tilde{\mathbf{x}})$ in Bob. In prior sections, we have already considered sources of ineffability up to $H_\phi(\mathbf{s} | \mathbf{m})$ and $K(\mathbf{s} | \mathbf{m}, p_\phi)$ in this chain. 
What remains is to identify additional sources of ineffability after the message is transmitted.
In this section we use the Kolmogorov formalism, since we assume the parameters $\phi$ of Alice's brain are not available to Bob. 

\subsubsection{A blank-slate listener}
\label{sec:blank_slate}

Before considering the case in which Bob is a typical human listener, we begin with a discussion of ineffability when Bob is a blank-slate (setting $\tilde{\phi} = \emptyset$, $\tilde{x} = \emptyset$, $\tilde{s} = \emptyset$, $\tilde{a} = \emptyset$, where $\emptyset$ denotes the null value). 
In this case the chain of communication ends at $\mathbf{m}$, thus a quantity of interest is the ineffability $K(\mathbf{s} | \mathbf{m})$ (without assuming access to Alice's cognitive parameters $\phi$, as we did in \cref{sec:message_info_loss}). 
Intuitively what this quantity refers to is the \emph{intrinsic} ineffability of an experience given its message, without conditioning on extra information such as cognitive parameters $\phi$ or $\tilde{\phi}$.
Taking an expectation to express average ineffability of conscious experience $\mathbf{s}$, we have $\mathbb{E}_{p_\phi(\mathbf{s} | \mathbf{m})} K(\mathbf{s} | \mathbf{m}) \geq \mathbb{E}_{p_\phi(\mathbf{s} | \mathbf{m})} K(\mathbf{s} | \mathbf{m}, p_\phi)$ trivially since conditioning on more information cannot increase the length of the shortest program that outputs $\mathbf{s}$, but it is important to note that one would additionally expect the reduction to be significant, i.e., $\mathbb{E}_{p_\phi(\mathbf{s} | \mathbf{m})} K(\mathbf{s} | \mathbf{m}) \gg \mathbb{E}_{p_\phi(\mathbf{s} | \mathbf{m})} K(\mathbf{s} | \mathbf{m}, p_\phi)$. 
This is because under Shannon's noiseless coding theorem, knowledge of Alice's state distribution $p_\phi$ reduces the problem of describing $\mathbf{s}$ in the general space of high-dimensional vectors to the problem of describing its index amongst the set of all possible conscious experiences associated with $\mathbf{m}$ for a brain parameterized by $\phi$.

The inequality $\mathbb{E}_{p_\phi(\mathbf{s}| \mathbf{m})} K(\mathbf{s} | \mathbf{m}) \gg \mathbb{E}_{p_\phi(\mathbf{s}| \mathbf{m})} K(\mathbf{s} | \mathbf{m}, p_\phi)$ relates to an observation at the core of the philosophical debate on ineffability: our descriptions of our experiences never seem to come close to capturing their full richness.
The gap is so significant that it has at times led some philosophers, scientists, and laypersons to the dualistic conclusion that conscious experiences are intrinsically indescribable, such that there is something more to their content than physically-embodied information encoded in neural activity.
Using our model, we argue that these intuitions do not necessarily imply a non-physical basis for conscious experience but may be explained by physically grounded and significant information loss that is a natural consequence of computational processing between the cognitive states underlying our experiences and the linguistic messages that we use to express them.

The increase in ineffability from not conditioning on $p_\phi$ also applies to the problem of describing the attractor state, i.e., $\mathbb{E}_{p_\phi(\mathbf{a} | \mathbf{m})} 
 K(\mathbf{a} | \mathbf{m}) \gg $ \linebreak $\mathbb{E}_{p_\phi(\mathbf{a} | \mathbf{m})} 
 K(\mathbf{a} | \mathbf{m}, p_\phi)$, 
 due to $\mathbf{a}$ being a high dimensional vector that represents the output of working memory and $\mathbf{m}$ being a relatively low dimensional vector representing a sentence.
 Note that ineffability of the attractor imposes a lower bound on the ineffability of the conscious experience under mild assumptions, thus if the former is large, so is the latter (Box \ref{math:ceiling}). 
While the representation of $\mathbf{m}$ is shared amongst individuals who speak the same language, the representation of $\mathbf{a}$ is unique to communicator Alice. 
Therefore, under the Kolmogorov formalism, there is complexity or information content in $\mathbf{a}$ that requires adopting Alice's representation space to reconstruct.

An analogy can be made with word symbols and word embeddings (or representation vectors) in deep learning models of natural language, as initially proposed by~\citet{bengio2000neural} and the earlier ideas on distributed representations of symbols~\citep{hinton1986distributed}. Essentially, every word in the system is associated with an arbitrary unique integer (the symbol) as well as a learnable vector (the embedding). As shown by ~\citet{bengio2000neural}, word embeddings can be used to represent semantics in a shared space, and can therefore help a model generalize to new sentences from training data comprising only a small subset of all possible sentences. Importantly, because the word symbols are arbitrary, they contain no information about the embeddings.
In a similar vein, when communicating using a message that simply conveys an attractor using a symbolic description $\mathbf{m} = f_{\phi}^M(\mathbf{a})$, we lose the rich representation of $\mathbf{a}$ that provides information on Alice's subjective experience.

The significant magnitude of $\mathbb{E}_{p_\phi(\mathbf{s} | \mathbf{m})} K(\mathbf{s} | \mathbf{m})$ and $\mathbb{E}_{p_\phi(\mathbf{a} | \mathbf{m})} K(\mathbf{a} | \mathbf{m})$ captures the blank-slate or tabula rasa case of the problem of ineffability: without assuming knowledge of the parameters of Alice's brain, experiences are highly ineffable using low dimensional descriptions such as typical verbal messages. Nonetheless, $K(\mathbf{s} | \mathbf{m}) \leq K(\mathbf{s}) < \infty$; our experiences are describable \emph{in principle}, even to a blank slate observer where no additional information is assumed.
Using a numerical scale to quantify ineffability allows us to convey the dual sense in which our experiences are, to varying degrees, both communicable and ineffable. 

\begin{mathbox}[label=math:ceiling]{Triangle inequalities for Kolmogorov complexity}
We have that $K(\mathbf{a}| \mathbf{m^*}) \stackrel{+}{<} K(\mathbf{a}| \mathbf{s^*}) + K(\mathbf{s}| \mathbf{m^*}) $ \citep[Theorem 4.1]{grunwald2004shannon}
where $\mathbf{m^*}$ is the shortest prefix program that outputs $\mathbf{m}$ and halts, and likewise for the other variables.
Thus $K(\mathbf{a}| \mathbf{m^*}) - K(\mathbf{s}| \mathbf{m^*}) \stackrel{+}{<} K(\mathbf{a}| \mathbf{s^*})$. 
From a similar application of the triangle inequality, we have $K(\mathbf{s}| \mathbf{m^*}) - K(\mathbf{a}| \mathbf{m^*}) \stackrel{+}{<} K(\mathbf{s}| \mathbf{a^*})$. 
Assuming the complexity of conscious experience is at least as great as the complexity of the working memory attractor, $K(\mathbf{s}) \geq K(\mathbf{a})$, we obtain 
$K(\mathbf{s}| \mathbf{a^*}) \geq K(\mathbf{a}| \mathbf{s^*})$ from $I(\mathbf{a} : \mathbf{s}) = K(\mathbf{a}) - K(\mathbf{a} | \mathbf{s^*}) =  K(\mathbf{s}) - K(\mathbf{s} | \mathbf{a^*})$. Therefore we have that the ceiling on relative ineffability of conscious experience $\mathbf{s}$ is equal or higher than for working memory attractor $\mathbf{a}$. \\


\end{mathbox}

\subsubsection{A typical listener} \label{s:interpersonal}




\paragraph{Cognitive similarity and effability.} 
In a realistic communication scenario, the cognitive parameters of listener Bob $\tilde{\phi}$ are given by a high-dimensional vector that provides information about Alice's parameters $\phi$ within the generic space of high-dimensional vectors, due to  shared physical environment (including cultural experience) and shared evolutionary background, and thus may be used to reduce the description length of $p_\phi$.
Trivially, we have that the expected ineffability of Alice's conscious experience can only improve by conditioning on Bob's parameters $\mathbb{E}_{p_\phi(\mathbf{s} | \mathbf{m})}  K(\mathbf{s} | \mathbf{m}) \geq \mathbb{E}_{p_\phi(\mathbf{s} | \mathbf{m})} K(\mathbf{s} | \mathbf{m}, p_{\tilde{\phi}})$.
However, we also obtain that a ceiling on the disadvantage of using Bob's parameters compared to Alice's parameters scales with the difference between them (Box \ref{math:interp}).

\begin{mathbox}[label=math:interp]{Cognitive dissimilarity and ineffability}
From \citet[Theorem 2.10]{grunwald2004shannon} we obtain for given $\mathbf{m}, p_\phi, p_{\tilde{\phi}}$ that $0 \leq \mathbb{E}_{p_\phi(\mathbf{s} | \mathbf{m})} [ K(\mathbf{s} | \mathbf{m}, p_{\tilde{\phi}})] - H_\phi(S | \mathbf{m}) 
\leq K(p_\phi(\cdot | \mathbf{m}) | p_{\tilde{\phi}}, \mathbf{m}) + c \leq K(p_\phi | p_{\tilde{\phi}}) + c $ 
where $c$ is a constant,
and $ 0 \leq \mathbb{E}_{p_\phi(\mathbf{s} | \mathbf{m})} [K(\mathbf{s} | \mathbf{m}, p_{\phi}) ] - H_\phi(S | \mathbf{m}) \leq K(p_\phi(\cdot | \mathbf{m}) | p_{\phi}, \mathbf{m}) + c = \epsilon  + c$, where $\epsilon$ is the negligible descriptional complexity of $p_\phi(\cdot | \mathbf{m})$ given $p_{\phi}$. 
Note $H_\phi(S | \mathbf{m}) \geq H_\phi(S | \mathbf{m}, p_{\tilde{\phi}})$ where the underlying joint distribution includes the meta-distribution over $p_{\tilde{\phi}}$, and likewise $H_\phi(S | \mathbf{m}) \geq H_\phi(S | \mathbf{m}, p_{\phi})$. 
Then $\mathbb{E}_{p_\phi(\mathbf{s} | \mathbf{m})} [ K(\mathbf{s} | \mathbf{m}, p_{\tilde{\phi}})] \leq H_\phi(S | \mathbf{m}) + K(p_\phi | p_{\tilde{\phi}}) + c$ and $\mathbb{E}_{p_\phi(\mathbf{s} | \mathbf{m})} [K(\mathbf{s} | \mathbf{m}, p_{\phi}) ] \leq H_\phi(S | \mathbf{m}) + \epsilon + c$.
The difference between upper bounds on ineffability is $ K(p_\phi | p_{\tilde{\phi}}) - \epsilon$.
\end{mathbox} 

The mismatch between Alice and Bob's parameters, which is formalized by $ K(p_\phi | p_{\tilde{\phi}})$ or the minimum number of bits required to encode a program that produces Alice's parameters from Bob's,
loosely corresponds to the difference between Bob and Alice's cognitive function (Box \ref{math:smoothness}), which depends on the extent to which they differ in genetic biases and lived experiences. 
This result supports the common intuition that our experiences are more effable or communicable to people who are similar to ourselves. 
It also resonates with the empirical observation of greater inter-brain synchronization in related individuals \citep{goldstein2018brain} and how the brain's anatomical structure (i.e. $\phi$ and $\tilde{\phi}$) affects the propensity to communicate at the inter-personal level \citep{dumas2012anatomical}.

Consider a prototypical example of inter-personal ineffability, in which Bob has been blind from birth and Alice is attempting to convey her experience of seeing the color red. In this case, Bob's brain might be so different from Alice's that the distance between their cognitive parameters $ K(p_\phi | p_{\tilde{\phi}})$ is sufficiently high that the benefit of conditioning on his own parameters is negligible. In other words, since $\mathbb{E}_{p_\phi(\mathbf{s} | \mathbf{m})} K(\mathbf{s} | \mathbf{m}, p_{\tilde{\phi}}) \leq K(p_\phi | p_{\tilde{\phi}}) + c + H_\phi(S | \mathbf{m})$ (Box \ref{math:interp}), if $K(p_\phi | p_{\tilde{\phi}})$ is large, then the ceiling on $\mathbb{E}_{p_\phi(\mathbf{s} | \mathbf{m})} K(\mathbf{s} | \mathbf{m}, p_{\tilde{\phi}})$, the ineffability of Alice's conscious experience given the message from Bob's perspective, is also large. 
Intuitively, when $K(p_\phi | p_{\tilde{\phi}})$ is small, the information required to communicate the functions $f_\phi^M$, $f_\phi^A$ and $f_\phi^S$ in order to reconstruct $\mathbf{s}$ from $\mathbf{m}$ is offloaded to $p_{\tilde{\phi}}$, which is given, thus reducing a ceiling on expected program length $\mathbb{E}_{p_\phi(\mathbf{s} | \mathbf{m})} K(\mathbf{s} | \mathbf{m}, p_{\tilde{\phi}})$. 


The cognitive dissimilarity factor $K(p_\phi | p_{\tilde{\phi}})$ is also implicated in Frank Jackson's famous thought experiment, color scientist Mary who has lived her whole life in an entirely black and white room and has learned exhaustive knowledge about the process of color perception, but nonetheless possesses a brain that is incapable of understanding the experience of color (i.e., she does not know what it is like to see red) \citep{jackson1986, Alter2006-ALTPCA}. Since her knowledge is exhaustive, she knows everything that anyone could possibly tell her about the experience of seeing something red. 
Jackson argues that when she finally sees something red, she nevertheless learns something new (``what it is like to see red''). It has been argued that since she already knew all of the physical facts, what she learned must have been a non-physical fact \citep{jackson1986, chalmers2010character}. Many philosophers have responded to this argument, developing different conceptions of how what Mary learns might be physical after all \citep{Alter2006-ALTPCA}. 
Our model can be understood as offering support to the physicalist account. It highlights how the ineffability $\mathbb{E}_{p_\phi(\mathbf{s}| \mathbf{m})} K(\mathbf{s} | \mathbf{m}, p_{\tilde{\phi}})$ of Alice describing her experience of color to Mary (who is playing the role of Bob), may be explained in part by the difference in their cognitive function. 
In other words, the ability to empathize with another person from a verbal report of their experience is aided by cognitive similarity or ease of reconstructing their cognitive function based on knowledge of one's own cognitive function, but simply memorizing a description of how the brain behaves in response to color does not imply one's brain is capable of responding in that manner upon being exposed to it or its reference (i.e., hearing the word ``red''), and it is similarity in cognitive behavior that is implicated in $K(p_\phi | p_{\tilde{\phi}})$.

The result in Box \ref{math:interp} states that high ineffability of Alice’s experience of color to Mary  implies high cognitive dissimilarity between Alice and Mary. Cognitive dissimilarity is not equivalent to knowledge inadequacy; knowing how brain should respond does not imply being able to execute such a response.
The view that Mary learns different cognitive behavior upon exposure to the color red is closest to the interpretation that she acquires a new ability \citep{Lewis1990-LEWWET}, as opposed to a new mode of presentation \citep{Loar1990-LOAPS}, a new relation of acquaintance \citep{Conee1985-CONPAP} or a reminder of something that in principle she must have had access to all along
\citep{Dennet2006-DENWRK, Rabin2011-RABCMA}. 



\begin{mathbox}[label=math:smoothness]{Difference in functionality and difference in parameters}
For a scalar valued function $h$ with bounded gradient magnitude, we have $h(\mathbf{x}, \tilde{\theta}) = h(\mathbf{x}, \theta) + (\tilde{\theta} - \theta)^\intercal \nabla_\theta h(\mathbf{x}, \theta) + \mathcal{O}(\lVert \tilde{\theta} - \theta \rVert^2) \leq h(\mathbf{x}, \theta) + \lVert \tilde{\theta} - \theta \rVert \lVert \nabla_\theta h(\mathbf{x}, \theta) \rVert + \mathcal{O}(\lVert \tilde{\theta} - \theta \rVert^2) $ by the Taylor expansion.
Assuming first order gradients are bounded by positive constant $C$, then we have $| h(\mathbf{x}, \tilde{\theta}) - h(\mathbf{x}, \theta) | \leq C \lVert \tilde{\theta} - \theta \rVert  + \mathcal{O}(\lVert \tilde{\theta} - \theta \rVert^2)$, i.e. an upper bound on the mismatch in functional output given parameterizations $\theta$ and $\tilde{\theta}$ scales with the Euclidian distance between them.
\end{mathbox}

\paragraph{Theory of mind.}

Evolution has optimized human beings to be skilled at inferring the thoughts of others, an ability termed ``Theory of Mind'' \citep{premack1978does,graziano2011human,graziano2015attention,kelly2014attributing}.
In our model, there is a link between theory of mind and ineffability. If cognitive functions $f_{\tilde{\phi}}^X$ and $f_{\tilde{\phi}}^S$ that produce Bob's conscious experience $\tilde{\mathbf{s}}$ are optimized for decoding $\mathbf{m}$ into Alice's conscious experience $\mathbf{s}$, then ineffability is reduced compared to reconstructing Alice's conscious experience from the raw message, $K(\mathbf{s} | \mathbf{m}, \tilde{\mathbf{\phi}}) \geq K(\mathbf{s} | \tilde{\mathbf{s}}, \tilde{\mathbf{\phi}})$, 
because part of the computation of reconstructing $\mathbf{s}$ is executed during inference of $\tilde{\mathbf{s}}$, meaning that the smallest program from $\tilde{\mathbf{s}} $ and $\tilde{\phi}$ to $\mathbf{s}$ would make use of $\tilde{\mathbf{s}}$ to reduce its residual work, shortening the descriptive length of the program. In the extreme case, if $K(\mathbf{s} | \tilde{\mathbf{s}}, \tilde{\phi}) \stackrel{+}{=} 0$, then
by definition Bob's cognitive function is optimal for inferring Alice's conscious experience, since no material additional information is required to determine $\mathbf{s}$. 

In turn, if Alice's parameters $\phi$ contain information about Bob's cognitive function or parameters $\tilde{\phi}$, she is capable of producing her message $\mathbf{m}$ in a way that maximises effability and minimizes $K(\mathbf{s} | \tilde{\mathbf{s}}, \tilde{\phi})$, since her cognitive functionality, including verbal reporting function $f_\phi^M$, depend on $\phi$.


\paragraph{The grounding problem.}

Two individuals will generally understand the same word or sentence in different ways.
For example, if a social group generally associates cats with femininity and dogs with masculinity, these associations may be inverted for someone who has a male cat and female dog. 
A reasonable model for ineffability would account for such differences in their experiences, regardless of whether the individuals detect such inter-personal discrepancies in their conscious thoughts or verbally express such thoughts.
This is taken into account in 2 ways by our model. First, analogously to the case of $\mathbb{E}_{p_\phi(\mathbf{s} | \mathbf{m})} K(\mathbf{s} | \mathbf{m}, \tilde{\phi})$, the ineffability of Alice's conscious experience given Bob's conscious experience $\mathbb{E}_{p_\phi(\mathbf{s}  | \mathbf{m})} K(\mathbf{s} | \tilde{\mathbf{s}}, \tilde{\phi})$ pays a penalty that scales with $K(p_\phi | p_{\tilde{\phi}})$, which measures a mismatch between $p_{\tilde{\phi}}$ and $p_\phi$ where the latter includes all the parameters in Alice's computation graph, including those that parameterize functions on input data $D$. 
This grounds Alice's $\phi$ in a representation that is shared with Bob's $\tilde{\phi}$; intuitively, if Bob's parameters implement a function that operates differently on inputs than Alice's, they do not inform on the latter and the ceiling on ineffability is increased via $K(p_\phi | p_{\tilde{\phi}})$.
In other words, the objective meaning of $\mathbf{s}$ is largely determined by how $\phi$ relates $\mathbf{s}$ and the input $\mathbf{d}$: for Bob to understand $\mathbf{m}$ well requires him to know something about that relationship in Alice's brain, which is given by her parameters $\phi$.
Second, conscious experience $\mathbf{s}$ depends on $\phi$, which includes Alice's long-term knowledge, therefore $\mathbf{s}$ is capable of containing information about the associations Alice makes in the process of generating her thoughts, and thus the latter may also be included in the reconstruction target of $K(\mathbf{s} | \tilde{\mathbf{s}}, \tilde{\phi})$.

\subsection{Phenomenal and access consciousness}



Having provided an information theoretic dynamical systems perspective on richness and ineffability, we now turn explicitly to the question of whether rich phenomenal experience exists and why we self-report that it does. 
We first highlight ambiguities in the meaning of access before contrasting two hypotheses for explaining the report of phenomenal experience. 

\subsubsection{Effability, accessibility, reportability}

Notions such as ``accessible'', ``reportable'' and perhaps ``effable'' are somewhat ambiguous. A benefit of our framework is that it allows us to distinguish between (at least) three distinct notions in the vicinity.

First, as we have presented it above, the notion of ``effability'' refers to the ability to accurately describe one variable by another, which implies it can be formalized using mutual information (\cref{sec:richness_and_ineffability}).

Second, ``access'' is interpretable in two different ways. Direct access is the notion of a variable $X$ being a direct input to a function or process $g$, meaning $g$ is defined on variable $X$, whereas informational access is the notion of $g$'s input variable $A$ sharing mutual information with $X$, $I(A; X) > 0$, corresponding to $X$ being effable with respect to $A$.
A process that has no direct access to $X$ may still have access to its information via inputs; if $M = g(A)$, process $g$ has access to information about $X$ if $I(A; X) > 0$. Thus a variable may be effable with respect to the input and output variables of a process without being directly accessible to the process.

Third, while a reporting process is in general a process or transformation that outputs to another process, we stipulate that ``reporting process'' may be understood to refer specifically to those that output to processes outside the cortex, such as cortical processes that encode speech or motor movements. 
We may then say that a variable is directly (or informationally) reportable if it is directly (or informationally) accessible by a reporting process, where the report corresponds to the output of the reporting process. 

Note that so construed, we may dissociate the three notions. Variable $X$ is effable to variable $M$ if they share mutual information but may not be directly accessible to the function that produces $M$, or alternatively variable $X$ may be directly accessible by $g$ but not directly reportable, if $g$ does not output to processes outside of the cortex. These distinctions will be helpful in what follows.


\subsubsection{Existence and report of phenomenal experience}


According to the Global Workspace Theory, information from diverse brain regions corresponding to a variety of perceptual or cognitive processes is selected for inclusion in the contents of a centralized processing workspace associated with working memory that coordinates and communicates with multiple subsystems, resulting in a rich space of ``highly differentiated'' states with ``high complexity'' \citep{tononi1998consciousness}.

The features of this global workspace system make it suitable as a framework for an analysis of consciousness (i.e., phenomenal consciousness), even if we do not assume that only items in workspace are conscious. 
The features of the global workspace system also make it a suitable target for modelling in terms of attractor dynamics, since by their nature, states amplified and sustained in a central processing workspace are attractors. Thus, our model allows for the refinement of theses concerning the relationship of consciousness to the global workspace.

Global workspace models of consciousness \citep{dehaene2001towards} generally divide representations into three classes:
\begin{enumerate}
\item Those not computed by working memory processes (unconscious).
\item Those mobilized in the workspace via amplification and made accessible to downstream processing (conscious).
\item Those computed by working memory processes but not sufficiently amplified or attended to be released by the workspace. 
\end{enumerate}

The latter includes non-attractor transient states in an attractor model of working memory, and being rich and unreportable, are a clear candidate for the basis of phenomenal experience \citep{dehaene2001towards}. It is a point of debate between adherents of the global workspace framework, whether or not items from the third class are indeed conscious. Some say no \citep{naccache2018and, Cohen2016-COHWIT-3}, others say yes \citep{Prinz2012-PRITCB-2}. 

Working memory processes are represented by function $f^X_\phi$ in our model. By allowing $f_\phi^S$ to be abstract, our model only specifies that $S$ is a deterministic projection of $\phi$, $X$ and $A$, and therefore is compatible with both views. 
If one assumes that attractor states are included in the content of consciousness and that the physical basis of transient states and attractor states in working memory is the same (i.e. they are differentiated by duration of attentional amplification, not location of neural circuitry), it would be reasonable to believe that transient states are also included in conscious awareness. 
If this is the case, then transient states are rich states that are consciously experienced but not directly accessible or reportable by downstream processes, while being partially verbally effable because of shared information with attractors which are directly reportable.
In this paradigm, the fleeting nature of transient states impacts their direct reportability but not their inclusion in conscious experience. This is an attractive position partly because it takes phenomenology seriously---people report their conscious experience being much richer than they are able to articulate.

Regardless of whether transient states are included in the contents of consciousness, the attractor model for working memory suggests a second explanation for the self-report of phenomenal experience: an attractor state may encode information about its basin of attraction and thus information loss. 
For example, point attractor states may include dimensions whose values estimate the size of its local basin, which is a measure of the information loss when going from transient states in trajectories within that basin to the attractor state itself.
This posits that rich experience exists, whether inside or outside the delimitation of consciousness, and its properties - such as richness - would be reportable, even if the transient states that support them are not.
It is plausible that conscious awareness of abstract attributes of transient states such as richness would be advantageous, for instance when reasoning about one's uncertainty, including for the purpose of anticipating the listener's uncertainty when engaging in theory of mind to minimize ineffability (\cref{s:interpersonal}). 

Our model supports an interpretation for Sperling's experiments ~\citep{sperling1960information}, where subjects briefly exposed to a grid of characters were generally able to report character identities for \emph{any} prompted row (containing $\sim4$ characters)  but subsequently not other rows, in addition to being able to report that they experienced observing more characters. 
An account for this behavior is that upon receiving the prompt to report a specific row, working memory contents represented by attractor state $\mathbf{a}$ contained the identities of characters in the prompted row, a summary over the grid (e.g. the number of characters and their arrangement) and an estimate of the information lost by the summary, whilst information sufficient to discriminate all characters existed in the processing pipeline but in upstream sensory state $\mathbf{v}$, from which $\mathbf{x}$ and $\mathbf{a}$ were computed.
Subsequently, as attractor state $\mathbf{a}$ is directly accessible to verbal reporting process $f_\phi^M$, the characters in the prompted row, grid details at summary level, and the presence of information loss were directly reportable, and full grid details (identities of all characters) were not. 
The latter holds irrespective of where the distinction between conscious and unconscious is drawn, i.e. whether $\mathbf{x}$, which might have contained sufficient information from $\mathbf{v}$ to discriminate all characters, is considered conscious or not.


These arguments suggest that Block's distinction between phenomenal and access consciousness is not due to a categorical difference between fundamentally different kinds of processing \citep{block1995} but rather to a difference in the representational stage of the same information processing function \citep{dehaene2001towards}, and that the existence of a rich phenomenological experience that exceeds our reporting abilities \citep{sperling1960information} is both justifiable and veridically reportable. Unpacking the implications of the model is an important task for future work.


\section{Conclusion}
\label{sec:discussion}

This paper characterizes the rich and ineffable nature of conscious experience from an information theoretic perspective. 
It connects the ordinary notion of ineffability with mathematical formalisms of information loss, describing how the latter arises as a result of computation in cognitive processing, how it is implemented by an attractor model for working memory, and how it may be increased by the compressed nature of language as well as differences in the cognitive processing functions of individuals.




Attractor dynamics may be considered an attentional process: out of many, one or a few states are selected. 
This connects our work not only to Global Workspace Theory but more broadly to research in machine learning on attention mechanisms.
We generally observe that attention, e.g., as introduced in deep learning by~\citep{bahdanau2014neural}, may be used to name any function that incurs significant information loss and is present in both artificial and biological cognitive systems, where it is---at present---commonly modelled by the family of attention-based and transformer architectures \citep{bahdanau2014neural,devlin2018bert,khan2022transformers,chorowski2015attention} and dynamical systems \citep{khona2022attractor,rolls2007attractor} respectively.  

In this work we use a simple model to reason about emitter-receptor communication, where the past is conditioned on implicitly via parameters $\phi$ and stochasticity in dynamics. An alternative would be to model more complex communication patterns explicitly.
We have also not considered learning objectives for function parameters. Doing so would enable a discussion on the generalization benefits of the inductive bias~\citep{goyal2022} giving rise to this information loss: intuitively, how simpler representations support robustness \citep{mathis1994computational} and the successful extrapolation of behavior beyond previously seen inputs.
Information bottlenecks are a popular training regularizer in machine learning \citep{tishby2000information,alemi2016deep}, but are understudied in the context of biologically plausible models, despite generalization ability being a key difference between humans and current artificial learning systems.
Considering the benefits of information loss may allow us to understand ineffability more deeply; not just how it arises, but why. 



\section*{Acknowledgements}

The authors thank the following institutions for sources of funding:  the Canada CIFAR AI Chair Program, the Canada Research Chair Program, UNIQUE, IVADO, NSERC, Samsung, and the Quebec government.

\bibliography{references}

\end{document}